\documentclass[twoside,12pt]{article}

\usepackage{graphicx}
\usepackage{amsmath}
\usepackage{amssymb}
\usepackage{indentfirst}    
\usepackage{float}      
\usepackage{afterpage}      
\usepackage{url}

\long\def\symbolfootnote[#1]#2{\begingroup%
\def\thefootnote{\fnsymbol{footnote}}\footnote[#1]{#2}\endgroup}

\newcommand{\be}{\begin{equation}}
\newcommand{\ee}{\end{equation}}
\newcommand{\bea}{\begin{eqnarray}}
\newcommand{\eea}{\end{eqnarray}}

\def\s2w{\sin^2\theta_W}

\begin{document}

\title{\vspace{1cm} The SAMPLE Experiment and Weak Nucleon Structure}

\author{E.J.\ Beise$^1$, M.L.\ Pitt$^2$, and D.T.\ Spayde$^3$\\
$^1$ University of Maryland, College Park, MD 20742, USA\\
$^2$ Virginia Tech, Blacksburg, VA 24061, USA \\
$^3$ University of Illinois at Urbana-Champaign, Urbana, IL 61801, USA }

\maketitle

\begin{abstract}

One of the key elements to understanding the structure of the nucleon is the role of its quark-antiquark sea
in its ground state properties such as charge, mass, magnetism and spin. In the last decade, parity-violating
electron scattering has emerged as an important tool in this area, because of its ability to isolate the
contribution of strange quark-antiquark pairs to the nucleon's charge and magnetism. The SAMPLE experiment at
the MIT-Bates Laboratory, which has been focused on $\overline{s}s$ contributions to the proton's magnetic
moment, was the first of such experiments and its program has recently been completed. In this paper we give
an overview of some of the experimental aspects of parity-violating electron scattering, briefly review the
theoretical predictions for strange quark form factors, summarize the SAMPLE measurements, and place them in
context with the program of experiments being carried out at other electron scattering facilities such as
Jefferson Laboratory and the Mainz Microtron.

\end{abstract}


\section{Introduction}
\label{sec:intro}

Over the last decade, parity violating electron scattering has become
a unique tool for probing the contribution of the nucleon's sea of
quark-antiquark pairs to its ground state electromagnetic structure.
Improvements in experimental techniques, particularly in the
development and delivery of intense and high quality polarized
electron beams, have led to the ability to measure parity-violating
asymmetries at the level of parts per million with relatively good
precision.  A few measurements from the first of these challenging
experiments have now been completed, and several additional
experiments are planned or underway. While theoretical advances have
provided guidance as to the expected magnitude of sea quark
contributions, robust predictions do not yet exist. In the next
several years, a number of new measurements are expected to become
available over a range of momentum transfers. These data should begin
to constrain the various calculations and help identify both the
magnitude of $s$-quark contributions and the mechanism by which they
might be present.  In addition, there is some promise that lattice QCD
calculations, which have evolved significantly in the last several
years, may be able to provide predictions to confront the data. In
this paper we review the present state of experiment and theory with
a specific emphasis on the SAMPLE experiment, which was carried out at
the MIT-Bates Laboratory in the 1990's and for which final analysis of
the data has recently been completed~\cite{Ito03,Spa03}.

The outline of this paper is as follows: in this section we will give
an overview of parity-violating electron scattering and its role in
the determination of the quark structure of the nucleon's
electromagnetic form factors, particularly that of strange
quark-antiquark pairs. We will also review some of the theoretical
developments that have recently taken place.  After an overview of
some of the experimental issues, we provide an in-depth look at the
SAMPLE experiment at the MIT-Bates Laboratory in
Section~\ref{sec:sample-exp}.  In Section~\ref{sec:compare} we will
place the SAMPLE experiment in context with existing measurements from
the HAPPEX program at Jefferson Laboratory and the PVA4 program at the
MAMI facility in Mainz, Germany, as well as compare the various
results with theoretical predictions.  We will conclude with a
discussion of the future program at Jefferson Laboratory (G$^0$ and
HAPPEX) and at Mainz, and make some brief remarks about future
measurements that intend to use parity-violating electron scattering
for precision tests of the Standard Model.

\subsection{Weak Form Factors and the Role of Strange Quarks}
\label{sec:formfactors}

The formalism of the electroweak interaction between electrons and quarks
can be found in many textbooks and several recent reviews of
parity violating electron scattering~\cite{KuS01,BeH01,BeM01,Mus94}.
Here we will largely follow the notation used in~\cite{Mus94}.
At low momentum transfer, the interaction of an electron with a
nucleon is usually cast in these terms by
defining a nucleon current for which the quark substructure is encapsulated
in form factors that are the observables of the interaction. The
invariant amplitudes associated with a single photon or $Z$-boson
exchange are
\begin{eqnarray}
\label{eq:amplitudes}
M_\gamma &=& \frac{4\pi\alpha}{q^2} e_l l^{\mu}J_{\mu}^{EM}
\nonumber \\
M_Z &=& -\frac{G_F}{2\sqrt{2}}\left(g_V^l l^{\mu} + g_A^l l^{\mu 5}\right)
\left(J_{\mu}^{NC}\right) \,
\end{eqnarray}
where $q$ is the four momentum transferred from the electron to the
nucleon. Throughout this paper we will use the standard experimental
notation for the four-vector momentum transfer squared, $Q^2 = -q^2 >
0$.  Note that the $Q^2$ dependence of the neutral weak propagator has
been suppressed since, at momentum transfers that are small compared
to the $Z$ and $W$ masses, the weak interaction is usually treated as
a contact interaction with a strength determined by the Fermi decay
constant $G_F = 1.16639(1)\times 10^{-5}$~GeV$^{-2}$~\cite{PDG03}.
The fermion electromagnetic and weak charges $e_l$, $g_V^l$, and
$g_A^l$ are listed in Table~\ref{tab:charges}.

\begin{table}
\begin{center}
\begin{tabular}{c|c|c|c}
\hline
Fermion & $e_l$ & $g_V^l$ & $g_A^l$ \\
\hline
 & & & \\
$\nu_e$, $\nu_\mu$, $\nu_\tau$ & 0 & 1 & $-1$ \\
$e$, $\mu$, $\tau$ & $-1$ & $-1+4\s2w$ & 1 \\
$u$, $c$, $t$ & $\frac{2}{3}$ & $1-\frac{8}{3}\s2w$ & $-1$ \\
$d$, $s$, $b$ & $-\frac{1}{3}$ & $-1+\frac{4}{3}\s2w$ & 1 \\
 & & & \\
\hline
\end{tabular}
\caption{Standard Model values for the elementary electromagnetic
and weak charges of the fermions.}
\label{tab:charges}
\end{center}
\end{table}

Because the electron is considered to be a pointlike particle with no
structure, the leptonic currents $l^{\mu}$ and $l^{\mu 5}$ are simply
$\overline{u}\gamma^\mu u$ and $\overline{u}\gamma^\mu\gamma_5 u$,
respectively. The nucleon currents, on the other hand, are expressed
in terms of hadron form factors sandwiched between nucleon spinors,
\begin{eqnarray}
J_{\mu}^{EM} \equiv
\langle N\vert\hat{J}_{\mu}^{EM}\vert N\rangle
&=& {\overline U}\left[ \gamma_{\mu} F_1^{\gamma ,N}(q^2)
+ i\sigma_{\mu\nu}q^{\nu}\frac{F_2^{\gamma ,N}}{2M}\right]U \nonumber \\
J_{\mu}^{NC} \equiv
\langle N\vert\hat{J}_{\mu}^{NC} \vert N\rangle
&=& {\overline U}\left[ \gamma_{\mu} F_1^{Z,N}(q^2)
+ i\sigma_{\mu\nu}q^{\nu}\frac{F_2^{Z,N}}{2M} +
\gamma_{\mu}\gamma_5 G_A^{Z,N} + \frac{q_{\mu}}{M}\gamma_5 G_P\right] U \, .
\end{eqnarray}
For completeness, we have included the pseudoscalar form factor
$G_P$ in this expression, however we will otherwise neglect it
since it does not contribute to parity-violating
electron scattering. The form factors are the observables in elastic
electron scattering from a nucleon target. $F_1$ for the proton(neutron)
is normalized to 1(0) at zero momentum transfer, and $F_2$ to the
anomalous part of the magnetic moments. At relatively low momentum
transfer, the Dirac and Pauli form factors $F_1$ and $F_2$ are commonly
expressed as the Sachs form factors $G_E$ and $G_M$,
\begin{equation}
G_E = F_1 + \frac{Q^2}{4M^2}F_2\, , \;
G_M = F_1 + F_2 \, ,
\end{equation}
resulting in the normalization of $G_E^\gamma$ to 1(0) for the proton
(neutron), and of $G_M^\gamma$ to the appropriate magnetic moment.  An
equivalent definition can be applied to the vector weak form factors
$F_{1,2}^{Z,N}$.

The form factors can be related to their quark substructure by
expressing them as a sum over contributions from each quark
flavor. Traditionally, the contributions from the heavy quarks
($c$,$t$, and $b$) are neglected so that the form factors are written
as contributions from $u$, $d$, and $s$ quarks, where each
contribution is weighted by the appropriate quark charge from
Table~\ref{tab:charges}.  The proton's Sachs electromagnetic and
vector neutral weak form factors become
\begin{eqnarray}
G_{E,M}^{\gamma} &=& \frac{2}{3} G_{E,M}^u - \frac{1}{3}
\left(G_{E,M}^d + G_{E,M}^s\right) \nonumber \\
G_{E,M}^Z &=& \left(1-\frac{8}{3}\s2w\right)G_{E,M}^u
+ \left(-1+\frac{4}{3}\s2w\right)\left(G_{E,M}^d + G_{E,M}^s\right) \, ,
\end{eqnarray}
where, because the underlying vector current is the
same for the photon and the $Z$-boson once the charges are factored out,
these two expressions are just
two different linear combinations of the same quark form factors.

If one next makes the assumption of charge symmetry
in the nucleon, one can consider the neutron electromagnetic form factors as
a third observable, and uniquely identify the three light quark
contributions to the nucleon's vector current. The assumption of charge
symmetry implies that the wave function of the $u$ quarks
in the proton are identical to the $d$ quarks in the neutron,
or that $G_{E,M}^u$ and $G_{E,M}^d$ in the proton are simply
interchanged in the neutron. With a small amount of algebra the
proton's weak form factor can then be rewritten in terms of observable
proton and neutron $EM$ form factors along with a residual
contribution from strange quarks, allowing direct access to the
strange quark piece,
\begin{equation}
\label{eq:gemz}
G_{E,M}^Z = \left(1-4\s2w\right)G_{E,M}^{\gamma ,p} - G_{E,M}^{\gamma ,n}
- G_{E,M}^s  \, .
\end{equation}
The electroweak axial form factor of the nucleon can be similarly
deconstructed to reveal the contribution of strange quarks to nucleon
spin. In the lowest order limit of single $Z$-boson exchange, only the
isovector and SU(3) singlet contributions survive, resulting in
\begin{equation}
G_A^Z(Q^2) = -\tau_3 G_A(Q^2) + \Delta s
\end{equation}
where $\tau_3=+1(-1)$ for $p(n)$, $G_A(0) = -(g_A/g_V) = 1.2670 \pm
0.0035$~\cite{PDG03} as determined from $\beta$-decay experiments, and
the quantity $\Delta s$ is the strange quark contribution to nucleon
spin.  As will be discussed later, the higher order corrections to
$G_A^Z$ are expected to be significant.  The $Q^2$-dependence of $G_A$
has been characterized by a dipole form factor, $1/(1+Q^2/M_A^2)^2$,
with the dipole mass $M_A$ experimentally determined from neutrino
scattering and from pion electroproduction. The value of $M_A$ from
electroproduction, which was measured to be 1.069$\pm$0.016
GeV~\cite{Lie99}, must be corrected for contributions from higher
order processes in order to directly compare with results from
neutrino scattering~\cite{Ber02}, leading to a corrected value of
1.014$\pm$0.016~GeV. The value of $M_A$ from neutrino scattering has
also recently been reevaluated, by refitting the neutrino data using
updated nucleon electromagnetic form factors~\cite{Bud03}. This has
shifted $M_A$ from the PDG world average of
1.026$\pm$0.02~GeV~\cite{PDG03} to a new value of 1.001$\pm$0.020~GeV,
now in good agreement with the electroproduction data.

The fact that one can uniquely extract information about strange quark
effects in the nucleon from elastic neutral-current processes was
pointed out by Kaplan and Manohar in 1988~\cite{KaM88}.  Soon after,
McKeown~\cite{BMcK89} and Beck~\cite{Bec89} showed how $G_{E,M}^Z$
could be measured using parity-violating electron scattering, which
led to the program of experiments at MIT-Bates, Jefferson Laboratory,
and the Mainz Microtron that are the focus of this review.

The parity-violating component of the amplitude in
equation~\ref{eq:amplitudes} arises from the cross terms of the axial
and vector currents,
\begin{equation}
M_{PV} = -\frac{G_F}{2\sqrt{2}}\left(g_A^l l^{\mu 5}J^{NC}_{\mu} +
g_V^l l^{\mu}J^{NC}_{\mu 5}\right)\,
\end{equation}
where the nucleon current has now been separated into its vector and
axial vector components. In order to observe this PV amplitude
experimentally, one typically uses a longitudinally polarized beam and
an unpolarized target and looks at the relative difference in cross
section for the scattering as one flips the polarization of the
incident beam along or against its momentum direction, {\it i.e.}
between its right-handed and left-handed helicity states. This
difference is directly proportional to the interference between
$M_{\gamma}$ and $M_{PV}$. For elastic scattering from a spin-1/2
target such as a nucleon, the asymmetry can be written as
\begin{equation}
\label{eq:pvee}
A_{PV} = \frac{d\sigma_R - d\sigma_L}{d\sigma_R + d\sigma_L}\,
 = -\frac{G_FQ^2}{4\pi\alpha\sqrt{2}}
\frac{A_E + A_M + A_A}{\left[\varepsilon \left(G_M^\gamma\right)^2 +
\tau \left(G_M^\gamma\right)^2\right]}
\end{equation}
where
\begin{eqnarray}
A_E &=& \varepsilon G_E^Z(Q^2) G_E^\gamma(Q^2) \nonumber \\
A_M &=& \tau G_M^Z(Q^2) G_M^\gamma(Q^2) \nonumber \\
A_A &=& -\left(1-4\s2w\right)
\sqrt{\tau\left(1+\tau\right)\left(1-\varepsilon^2\right)} G_A^e(Q^2)
G_M^\gamma (Q^2)\, ,
\end{eqnarray}
and
\begin{equation}
\tau = \frac{Q^2}{4M_N^2}\; , \mathrm{and}\;
\varepsilon = \frac{1}{1+2\left(1+\tau\right)\tan^2\frac{\theta}{2}} \, .
\end{equation}

The $\gamma$-$Z$ interference is explicitly seen in these
expressions. Depending on the kinematics, one can tune an experiment
to be sensitive to the electric, magnetic, or axial form
factors. Forward angle experiments are typically sensitive to a
combination of $A_E$ and $A_M$, while backward angle experiments
determine a combination of $A_M$ and $A_A$.  Quasielastic scattering
from an isospin 0 target such as the deuteron can be used to enhance
$A_A$; this is discussed in more detail below.

\subsubsection*{Charge Symmetry Violation}

Isospin violation, or less restrictively charge symmetry breaking,
invalidates the assumption that the $u$-quark wave function in the
proton is identical to the $d$-quark wave function in the neutron and
would lead to an additional term in $G_{E,M}^Z$ that would be
indistinguishable from $G_{E,M}^s$. This violation arises from $u$ and
$d$ quark mass differences and from electromagnetic effects.
Dmitrasinovic and Pollock~\cite{Dmi95} and Miller~\cite{Mil98} both
calculated a charge symmetry breaking term within the context of a
non-relativistic constituent quark model. In such a model, the size of
the effect is driven by the ratio of the mass difference to the
constituent quark mass, about 1/70, and results in fractional
contributions to the isoscalar electric and magnetic form factors on
the order of 0.1\%. These models would thus predict that the precision
of the strange quark measurements would have to be better than this
level before charge symmetry breaking adds significant uncertainty to
their determination. Similar conclusions were reached by Lewis and
Mobed~\cite{Lew99}, who used a two-flavor heavy baryon chiral
perturbation theory (HB$\chi$PT) to investigate the effects of isospin
breaking.

\subsubsection*{Electroweak Radiative Corrections}

\begin{figure}
\begin{center}
\includegraphics[width=10cm]{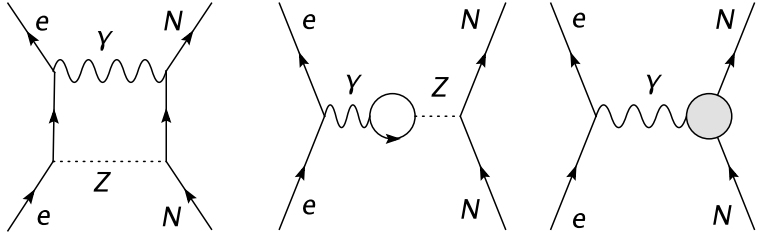}
\caption{Higher order corrections, such as $\gamma$-$Z$ box diagrams (a),
$\gamma$-$Z$ mixing corrections (b), and vertex corrections (c), that contribute to
the neutral current scattering amplitude. The third diagram can involve a
$Z$ exchange between two quarks in the nucleon. Both (b) and (c) are
classified as ``anapole'' corrections.}
\label{fig:gammaZ}
\end{center}
\end{figure}

Higher order diagrams, such as those depicted in
Figure~\ref{fig:gammaZ}, have been treated by several authors.  The
corrections to the vector weak form factors tend to be dominated by
terms involving a single quark and can be directly computed within the
context of the Standard Model. They have been calculated
at low energy (see references in~\cite{PDG03}), and at the low
values of momentum transfer relevant to the experiments described
here, the dependence on $Q^2$ is relatively weak. These contributions
have the same $(1-4\s2w )$ multiplier as the tree-level amplitudes and
thus are typically quite small. The one-quark axial corrections can
also be computed in a model-independent fashion with relatively small
uncertainty but they are substantial relative to the tree-level
contribution. Of more concern in the axial radiative corrections are
the potentially large contributions from diagrams involving two or
more quarks.  One class of such diagrams, referred to as ``anapole''
terms, involve an electromagnetic interaction with the scattered
electron but weak interactions within the target. The calculations of
weak radiative corrections most relevant to this paper are those of
Musolf and Holstein~\cite{Mus90} and of Zhu~{\it et
al.}~\cite{Zhu00}. In~\cite{Mus90}, the many-quark corrections are
modeled as effective parity-violating hadronic couplings within the
nucleon with the photon coupling to the nucleon via a meson loop. The
dominant source of uncertainty results from the uncertainties in the
hadronic couplings as determined by Desplanques, Donoghue and
Holstein~\cite{DDH80}.  Zhu~{\it et al.}~\cite{Zhu00} carried out a
similar analysis, but recast the hadronic couplings in a heavy baryon
chiral perturbation theory framework. All contributions through ${\cal
O}(1/\Lambda_\chi)$ were included, where $\Lambda_\chi=4\pi F_{\pi}$
is the scale of chiral-symmetry breaking. Potential contributions from
additional mesons, as well as couplings involving kaons, were also
included.  The terms involving kaon loops were generally found to be
small compared with those involving pions. As in the previous
analysis, it was found that the one-quark axial corrections dominate
the magnitude of the correction, whereas the multiquark terms dominate
its uncertainty.

Incorporating the higher order terms as weak radiative corrections
leads to the following modification of equation~\ref{eq:gemz}:
\begin{equation}
\label{eq:gemz_rad}
G_{E,M}^Z = \left(1-4\s2w\right)\left(1+R_V^p\right)G_{E,M}^{\gamma ,p}
- \left(1+R_V^n\right)G_{E,M}^{\gamma ,n}
- G_{E,M}^s  \,
\end{equation}
where the corrections to $G_{E,M}^s$ are left as implicit. The axial
form factor is also modified, and the notation changed to $G_A^e$,
in order to distinguish the form factor as seen by electron-scattering
from that seen by neutrino scattering where the higher order diagrams
involving a photon are absent, giving
\begin{equation}
G_A^e(Q^2) = -\tau_3(1+R_A^{T=1})G_A + \sqrt{3}R_A^{T=0}G_A^8 + \Delta
s \, .
\end{equation}
There is now explicitly a term proportional to an SU(3) isoscalar
octet form factor $G_A^8$, which at tree-level is zero.  The magnitude
of this form factor is estimated (it has not been directly measured)
using the ratio of axial vector to vector couplings in hyperon beta
decay, which, assuming SU(3) flavor symmetry, can be related to the
octet axial charge $a_8$ and to $F$ and $D$ coefficients~\cite{PDG03}.
\begin{equation}
G_A^8(0) = \frac{(3F-D)}{2\sqrt{3}} = \frac{1}{2}a_8 = \frac{\sqrt{3}}{2}(-0.25\pm0.05)
= 0.217\pm 0.043 \, .
\end{equation}
The $Q^2$ dependence is also not measured, but can be estimated using
the same dipole mass behavior as for the isovector axial form factor.
This results in a net isoscalar octet correction that we define as
$R_A^0 = \sqrt{3}R_A^{T=0}G_A^8(Q^2)$, which for $Q^2=0.1$~(GeV/c)$^2$
is $0.02\pm 0.04$. Table~\ref{tab:weakcor} contains the values of the
vector and axial vector weak radiative corrections used in computing
the parity-violating asymmetry relevant to the SAMPLE experiment.

\begin{table}
\begin{center}
\begin{tabular}{| c | c | c |}
\hline
correction & $T=0$ & $T=1$ \\
\hline
$R_V$ & $-0.0113$ & $-0.017\pm0.002$ \\
$R_A$ & $0.06\pm 0.14$ & $-0.23\pm0.24$ \\
\hline
\end{tabular}
\caption{Values of electroweak radiative corrections to parity-violating electron
scattering, as described in the text, computed in
the $\overline{MS}$ scheme. The vector corrections are taken
from~\protect{\cite{Mus94}}, and the axial corrections are taken
from~\protect{\cite{Zhu00}} after the 1-quark corrections are converted to
their $\overline{MS}$ values. Note that the nucleon vector corrections
$R_V^p = [(1-2\s2w)R_V^{(T=1)} - 2\s2w R_V^{(T=0)}]/(1-4\s2w)
= -0.054\pm0.033$
and $R_V^n = (1-2\s2w)R_V^{(T=1)} + 2\s2w R_V^{(T=0)}=-0.0143\pm0.0004$ are used
in the text.}
\label{tab:weakcor}
\end{center}
\end{table}

\subsubsection*{Sensitivity to Nucleon Electromagnetic Form Factors}

An important consideration in extracting the strange vector
form factors is the quality of information known
about the nucleon's electromagnetic structure.
An interesting consequence of the isospin structure of the
neutral weak interaction is that, as can be observed
in equation~\ref{eq:gemz}, determination of the strange quark form factors
of the {\it proton} requires good knowledge of the {\it neutron's}
electromagnetic properties.

Precise determination of $G_E^\gamma$ and $G_M^\gamma$ for
both the proton and neutron has been a topic of intense activity at
all of the electron scattering laboratories over the last decade and
significant progress has been made in recent years. The majority of
the information on the proton charge and magnetic form factors comes
from unpolarized electron scattering experiments. In the one-photon
exchange approximation, the $e$-$p$
elastic scattering cross section in the laboratory frame
can be written as
\begin{equation}
\label{eq:ep_xsect}
\frac{d\sigma}{d\Omega} = \left(\frac{\alpha^2 \cos^2\frac{\theta}{2}}
{4E^2\sin^4\frac{\theta}{2}}\right)\frac{E^\prime}{E}
\frac{1}{\varepsilon \left(1+\tau\right)}
\left[\varepsilon \left(G_E^\gamma \right)^2
+ \tau \left(G_M^\gamma \right)^2\right] \,
\end{equation}
and through measurements at various scattering angles one can separate
the electric and magnetic pieces~\cite{HaM84}.
Very recently, this ``Rosenbluth'' technique has
come under some degree of scrutiny because of new results from
polarization experiments that are in conflict with the results
from the unpolarized data at momentum transfers above 1 (GeV/c)$^2$.
While we will not attempt a detailed review of this topic,
some brief remarks are in order.

With a polarized beam and either a polarized target or detection of
the polarization of the recoiling nucleon it is possible to determine
directly either the ratio of $G_E^\gamma /G_M^\gamma$, or the product
$G_E^\gamma G_M^\gamma$ in a single measurement.  These techniques do
not require an absolute determination of the experimental cross
section and are thus less susceptible to experimental systematic
uncertainties related to knowledge of detector acceptance, absolute
beam flux, or detector efficiencies. They are also less sensitive to corrections
beyond the one-photon exchange approximation. With a longitudinally polarized
beam and detection of the recoil polarization, the ratio of
polarization components perpendicular and parallel to the nucleon's
momentum determines the ratio~\cite{Akh68,Dom69,Akh74,Arn81}
\begin{equation}
\label{eq:recpol}
\frac{G_E^\gamma}{G_M^\gamma}=-\frac{P_T}{P_L}
\frac{E_e+E_e^\prime}{2M_N}\tan\frac{\theta_e}{2} \, ,
\end{equation}
where $(E_e,E_e^\prime,\theta_e)$ are the incident and scattered electron
energy and electron laboratory scattering angle, respectively.

Recently, data have been taken at Jefferson Laboratory using the
recoil polarization technique for $e$-$p$
scattering~\cite{Jon00,Gay02,Pun03} in which a monotonic decrease of
$\frac{G_E^\gamma}{G_M^\gamma}$ with increasing $Q^2$ was found.  This
is in contrast to a global analysis of the world's set
of Rosenbluth data that seemed to
indicate a more constant ratio~\cite{Arr03}.  New cross section data
from Jefferson Laboratory are in good agreement with the older cross
section data~\cite{Chr03}.  Radiative corrections coming from
two-photon exchange introduce an additional $\varepsilon$-dependence
to the cross section that distort the relative contributions of
$G_E^p$ and $G_M^p$, complicating the Rosenbluth extraction at high
momentum transfer where the $\varepsilon$-dependent term is typically
only a small fraction of the cross section. Preliminary calculations
that do not include intermediate states of the nucleon indicate that
two-photon exchange can account for at least half of the discrepancy
between the two data sets~\cite{Gui03,Blu03}, as well as for differences
between electron-proton and positron-proton scattering~\cite{Arr03b}.
New experiments are underway that will provide additional data on the proton form factors
at high momentum transfer~\cite{Arr03a,Per03} and new measurements are
being proposed to explicitly measure two-photon exchange
processes~\cite{JPAC04}.  At the time of this writing, the discrepancy
between the results from the unpolarized cross section data and the
polarization measurements has yet to be completely resolved. At the
momentum transfers of interest to the discussion here (below 1
(GeV/c)$^2$), the data from the two methods agree within experimental
uncertainties and both of the proton's electromagnetic form factors
are taken to be known to 2\%~\cite{Arr03b}.  Polarized targets have
not been used in the past to determine proton EM form factors, but an
experiment is planned at the MIT-Bates Laboratory in the near
future~\cite{BLAST} using this technique.

It should also be noted that experiments with an unpolarized target and
transversely polarized beam are directly sensitive to the imaginary part
of the two-photon amplitudes. Experimentally one measured an azimuthal dependence
to the yield asymmetry.  Such measurements can typically be performed to
ppm-level precision in a few days in a parity-violation experiment setup.
While these imaginary components do not influence the radiative corrections that
enter the cross section measurements, their determination can provide
additional constraints on the theory. As will be discussed below, data
presently exist from the SAMPLE and PVA4 experiments and are expected from
the G0 experiment.

While the neutron's electromagnetic structure is not yet known to this
degree of precision, considerable progress has been made in recent
years.  In fact, it is the neutron form factors to which the parity
violation experiments that seek to extract strange quark information
are more sensitive because,
to lowest order, the neutron's vector neutral weak charge is large
relative to that of the proton.  The majority of the new measurements
of the neutron's electric and magnetic form factors have used
polarization techniques and, by necessity, light nuclear targets,
causing an additional level of complexity in extracting the form
factors.

With a polarized beam and a polarized target, the differential cross
section for elastic scattering has an additional spin-dependent
term $\Delta$,
\begin{equation}
\frac{d\sigma}{d\Omega} = \Sigma + h\Delta
\end{equation}
that depends on the electron beam helicity, $h$=$\pm$1.
The unpolarized differential cross section $\Sigma$ was defined
in eq.~(\ref{eq:ep_xsect}). By reversing the electron
beam helicity and measuring the asymmetry in the count rate for
a given target polarization, the spin-dependent term can be isolated,
through the spin-dependent yield asymmetry
\begin{eqnarray}
A &=& \frac{d\sigma^+ - d\sigma^-}{d\sigma^+ + d\sigma^-} \nonumber \\
 &=& \frac{\Delta}{\Sigma} = -\frac{2\tau v_{T^\prime}\cos\theta^*
\left(G_M^\gamma \right)^2 + 2\sqrt{2\tau\left(1+\tau\right)}
v_{TL^{\prime}}\sin\theta^*\cos\phi^*G_M^\gamma G_E^\gamma}
{\left(1+\tau \right)v_L\left(G_E^\gamma \right)^2
+2\tau v_T\left(G_M^\gamma \right)^2} \, ,
\end{eqnarray}
where $(\theta^*,\phi^*)$ define the direction of the target polarization
with respect to the three-momentum transfer vector ${\bf q}$ and the
$e$-$p$ scattering plane, and $v_i$ are kinematic factors.
Polarized deuterium and $^3$He targets have been used to extract
neutron form factors, where the use of polarization observables has
provided a significant reduction in the nuclear model dependence.  A
recent review of the various experimental techniques can be found
in~\cite{Gao03}. Here we simply highlight the particularly notable
features of the recent progress relevant to the determination of
strange quark form factors.

\begin{figure}
\begin{center}
\includegraphics[angle=90,width=10 cm]{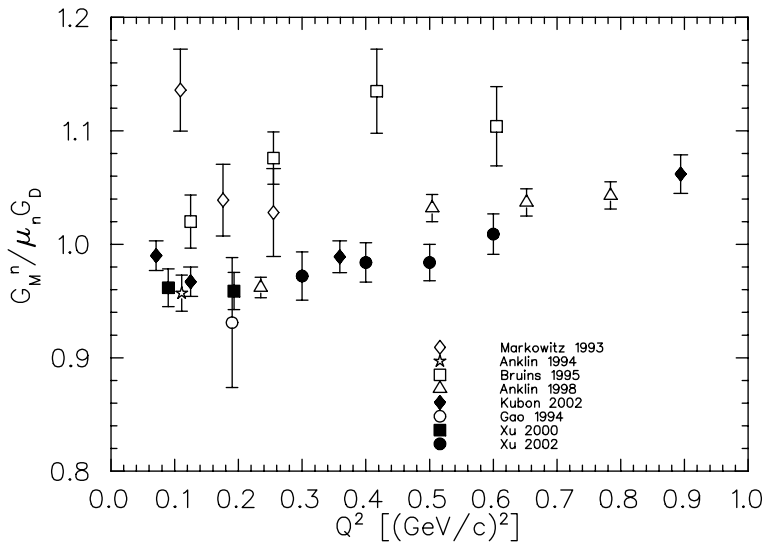}
\caption{Recent measurements of the neutron's magnetic form factor. Figure taken
from~\protect{\cite{Gao03}}. The two most recent data sets are
from~\protect{\cite{Xu03}} (solid circles) in which the asymmetry in
$\vec{^3\mathrm{He}}(\vec{e},e^\prime)$ was measured,
and~\protect{\cite{Kub02}}(solid diamonds) using the ratio of unpolarized
cross sections $d(e,e^\prime n)/(e,e^\prime p)$.}
\label{fig:gmn}
\end{center}
\end{figure}

A series of experiments was carried out at Jefferson Laboratory with a
polarized $^3$He target to determine $G_M^n$ up to a momentum transfer
of 1~(GeV/c)$^2$~\cite{Xu00,Xu03}. In these measurements, it was found
that below $Q^2\sim 0.3$~(GeV/c)$^2$ final state interaction effects
modify the asymmetry significantly, requiring the use of a
state-of-the-art three-nucleon wave function to relate the asymmetry
to the neutron form factor.  Above 0.3~(GeV/c)$^2$, final state
interactions are less important but relativistic corrections become
significant, so a PWIA calculation that includes relativistic
corrections was used to extract $G_M^n$. The data are shown in
Figure~\ref{fig:gmn} as the solid circles.  The
$\vec{^3\mathrm{He}}(\vec{e},e^\prime)$ asymmetry measurements can be
compared to another recent determination of $G_M^n$ using the ratio of
unpolarized cross sections $d(e,e^\prime n)/(e,e^\prime
p)$~\cite{Kub02}, shown as the solid diamonds in
Figure~\ref{fig:gmn}. Nuclear structure effects cancel to a large
extent in the ratio, and the dominant uncertainty in the measurement
comes from absolute knowledge of the neutron detection
efficiency. This has been a source of discrepancy between past
measurements~\cite{Bru95,Ank98}. In~\cite{Kub02} the neutron detection
efficiency was measured using tagged and monoenergetic neutron beams
with a claimed accuracy of 1\%, resulting in an uncertainty in $G_M^n$
of less than 2\%.  The agreement between these two most recent
determinations of $G_M^n$, using quite different experimental
techniques with very different theoretical uncertainties, is
remarkable. A new experiment using the cross section ratio technique,
has recently been completed at JLab, which will push the measurements
to higher momentum transfer~\cite{Bro94}. So while precise information
at high momentum transer is still pending, the low-momentum transfer
data have now reached a level of precision comparable to our knowledge
of $G_M^p$.

There has also been much recent progress on the neutron's electric
form factor.  Initial polarization measurements demonstrated the
feasibility of extracting $G_E^n$ in a somewhat model-dependent way,
but were limited by statistical precision. Schiavilla and
Sick~\cite{Sch01} showed recently that one can reduce the model
uncertainties in extracting $G_E^n$ from elastic $e$-$d$ scattering
data by using only the deuteron's quadrupole form factor, which has
been experimentally determined from the deuteron's tensor
polarization~\cite{Abb00} and is much less sensitive to
uncertain short-range two-body currents in the deuteron. Two recent
JLab experiments, one using a polarized target~\cite{War03} and the
other using recoil polarimetry~\cite{Mad03}, now provide the first
double polarization measurements at momentum transfers above 1
(GeV/c)$^2$. The measurements are in good agreement with each other,
and, combined with recent results from Mainz at lower momentum
transfer using $^3\vec{\mathrm{He}}$ and $\vec{d}$
targets~\cite{Ost99,Ber03}, have greatly improved our knowledge of
$G_E^n$.  The BLAST collaboration at MIT-Bates~\cite{BLAST} will take
data using both $^3\vec{\mathrm{He}}$ and $\vec{d}$ targets at low
momentum transfer over the next year, providing additional systematic
checks of the nuclear structure contributions.  It should be noted
that uncertainties in $G_E^n$ at the 10\% level will not significantly
impact the extraction of the strange quark form factors with the
proposed level of precision of the near term parity violation
experiments.

\subsection{Models of the Strange Vector Form Factors}
\label{sec:s_models}

Over the last decade a number of different models have been
used as a context in which to estimate the magnitude of
the strange quark vector form factors. A recent theoretical
review of the origins of many of the models can be
found in~\cite{BeH01}: here we will simply give a qualitative
picture and review some of the predictions. The nucleon has
no net strangeness, so $G_E^s(0)$=0, but neither the sign nor
the magnitude of $G_M^s(0)$, referred to in the literature
as $\mu_s$, is yet known. It has become conventional
to estimate the leading $Q^2$-dependence of $G_M^s$ and
$G_E^s$, which are defined by a mean-square radius derived
from the slopes at $Q^2=0$. We use the following definitions,
used by Jaffe~\cite{Jaf89} and Musolf {\it et al.}~\cite{Mus94}\footnote[2]
{One also finds in the literature the dimensionless Dirac strangeness
radius, $\rho_D = -\frac{2}{3}M_N^2r_s^2 = \rho_s + \mu_s$, and the
dimensionful Sachs radius $r_s^2(S) = -\frac{3}{2M_N^2}\rho_s = -0.066\rho_s$
in fm$^2$.}
\begin{eqnarray}
r_s^2 &\equiv& -6 \frac{dF_1^s}{dQ^2}|_{Q^2=0}\, , \nonumber \\
\mu_s &=& G_M^s(0) = F_2^s(0) \nonumber \\
\rho_s &\equiv& \frac{dG_E^s}{d\tau}|_{Q^2=0}\,
= -\frac{2}{3}M_N^2r_s^2 - \mu_s \, .
\end{eqnarray}


\subsubsection*{``Poles'' and Dispersion Relations}
\label{sec:poles}

One of the earliest estimates of the magnitude of strange quark vector
form factors came from Jaffe~\cite{Jaf89}, who used a vector meson
dominance (VMD) model and a dispersion analysis comparable to that
used by H\"ohler~\cite{Hoh76} for the electromagnetic form factors.
In the VMD framework, the $\sim\frac{1}{Q^4}$ ``dipole'' behavior of
the nucleon form factors, ultimately arising from the quark-antiquark
sea, is generated by intermediate state vector meson resonances such
as the $\phi$ and $\omega$ and a higher mass meson that incorporates
all additional unknown contributions.  H\"ohler had noted that, in
such a 3-pole fit to the form factors, a strong coupling to a meson
with a mass close to the predominantly $s\overline{s}$ $\phi$ meson
was required to get the dipole behavior. Using H\"ohler's fit to the
isoscalar form factors, the mixing of the physical $\omega$, $\phi$
from their pure $q\overline{q}$ states
($(u\overline{u}+d\overline{d})/\sqrt{2}$, and $\overline{s}s$,
respectively) as determined by radiative $\phi$ decay, and assumptions
about the asymptotic behavior of the form factors, the behavior of
$F_1^s$ and $F_2^s$ at low momentum was deduced. The result was a
value of $\mu_s\sim -0.3$ and $r_s^2\sim 0.15$~fm$^2$.

This analysis has been refined by several authors in various
ways. Mergell, Meissner, and Drechsel~\cite{Mer96} updated H\"ohler's
analysis including newer, more precise, form factor data and
Hammer~{\it et al.}~\cite{Ham96} used the updated fit to revisit
Jaffe's analysis.  They found $\mu_s$=$-0.24\pm0.03$, and
$r_s^2=(0.21\pm 0.03)$~fm$^2$.  Forkel~\cite{For97} also refit the
data, finding similar results. He also revisited the asymptotic
behavior used by Jaffe, requiring a better match to that predicted by
quark counting rules at large momentum transfer. To change the
asymptotic behavior he introduced additional poles, and thus
additional fit parameters, but generally found this reduced both
$r_s^2$ and $\mu_s$ by about a factor of 2.

Finally, in a series of papers, Hammer, Ramsey-Musolf, and
collaborators~\cite{MHD97,MuH98,Ham99a,Ham99b} took the dispersion
relation analysis beyond the context of simple vector meson dominance
and also studied possible continuum contributions, particularly that
of $K\overline{K}$. They found that in a combined fit to the isoscalar
(nonstrange) form factors most of the contribution to $F_2^{(I=0)}$
from the $\phi$-meson is taken up by the $K\overline{K}$ continuum. In
the case of $F_1^{(I=0)}$, some strength from the $\phi$ remains,
although good fits are also obtained if the $\phi$ strength is
replaced with a $\rho-\pi$ continuum strength, leaving the role of the
$\phi$-meson in the structure of $F_1$ more ambiguous. When they use
the results of their fit to the EM form factors to extract the
low-$Q^2$ behavior of the strange form factors, they get a comparable
value for $\mu_s$, $-0.28$, as had been obtained in~\cite{Jaf89}
and~\cite{Ham96}, but a significantly larger value for the electric
strangeness radius $r_s^2$ =0.42~fm$^2$.

\subsubsection*{Kaon Loops}
\label{sec:loops}

Another approach to estimating strange quark effects has been the
``kaon loop'' picture, where the origin of the spatial separation of
$\overline{s}$ and $s$ comes from a $\Lambda$-$K$ component to the
proton's wave function. Typically these models predict a negative
value for $r_s^2$, consistent with the picture of a $\Lambda$ core
surrounded by a kaon cloud.\footnote[3]{The convention established
in~\protect{\cite{Jaf89}} and followed by most subsequent calculations
is that a positive strangeness radius corresponds to an $s$ at larger
radii on average.} Riska and collaborators~\cite{Han00a} have also
pointed out that this simple picture combined with helicity flip
arguments would lead to a negative value of $\mu_s$.


A kaon cloud contribution within the context of a constituent quark model
was considered by Koepf, Henley and Pollock~\cite{Koe92}. They
considered only contributions from the lightest kaons, and found that
the use of pointlike quarks results in significant $s$-quark contributions
but also in poor agreement with the nucleon's measured
electromagnetic properties. When they instead used a nucleon
model with some spatial extent, such as a cloudy bag model, the result
was very little contribution from strange quarks, an order of magnitude
smaller than the estimate from the vector meson dominance models.
However, their results are very sensitive to the choice of bag radius.
Musolf and Burkardt~\cite{Bur94} also looked predominantly at the
contribution from ground state kaons but used hadron scattering data
to constrain the form factors used at the meson-nucleon vertices.
They also included seagull graphs required for gauge invariance. Adding
these terms resulted in a value of $\mu_s$ again close to that of the
vector meson dominance approach with little change to $r_s^2$.
However, Geiger and Isgur showed that summing over a complete
set of strange meson-baryon intermediate states results in cancellations
that again reduce $\mu_s$ to a very small (and in their case positive)
value~\cite{Gei97}.


\subsubsection*{Chiral Perturbation Theory}
\label{sec:chipt}

Recent efforts have been made to use chiral perturbation theory
($\chi$PT) to compute the leading $Q^2$ behavior of $G_E^s$ and
$G_M^s$.  At low energies, $\chi$PT has been enormously successful in
describing a wide variety of observables and several authors have
pursued the heavy-baryon version of $\chi$PT (HB$\chi$PT) to study
nucleon form factors.  In~\cite{Hem98} Hemmert and collaborators
showed that to one-loop order (or ${\cal O}(p^3)$ where $p$ is the
chiral expansion parameter), the $Q^2$ dependence of $G_M^s$ is free of
unknown parameters, while $\mu_s$ requires knowledge of two
unknown counterterms.  Ramsey-Musolf and Ito~\cite{Ito97} pointed out
that higher order terms that could not be constrained by experiment
would likely be as important in flavor singlet operators such as
$\overline{s}\gamma_\mu s$. Nonetheless, Hemmert and collaborators
combined their calculated $Q^2$-dependence for $G_M^s$ with the
earliest results from SAMPLE~\cite{Mue97} and HAPPEX~\cite{Ani99} to
predict the behavior of $G_M^s$ and $G_E^s$ over a broad range.
They found that the two form factors must have opposite sign and
result in $\rho_s=-0.8\pm 1.4$. The large uncertainty is driven
primarily by extrapolation of the HAPPEX data back to
$Q^2=0$. Recently, Hammer~{\it et al.}~\cite{Ham03} computed the
$Q^2$-dependence of $G_M^s$ to next order in $\chi$PT and found the
${\cal O}(p^4)$ to largely cancel the the ${\cal O}(p^3)$ terms,
leaving the leading behavior of $G_M^s$ to be determined by additional
unknown constants such that even the sign of $\frac{dG_M^s}{dQ^2}$ is
not known. Since experiments cannot directly measure $G_M^s(0)$, some
form of extrapolation will be required to compare directly to
calculations at this static limit. An experimental study of the $Q^2$
behavior of $G_M^s(Q^2)$ is thus required. The authors of
Ref.~\cite{Ham03} formulate the extrapolation problem by writing
$\mu_s$ as
\begin{equation}
\mu_s = G_M^s(Q^2) - 0.13 b_s^r
\end{equation}
where $b_s^r$ is the relevant unknown constant, giving a reasonable range from
dimensional arguments of $\vert b_s^r\vert\leq 1$.


\subsubsection*{Lattice QCD}
\label{sec:lattice}

Ultimately, lattice QCD techniques will lead to the most rigorous
theoretical prediction for the nucleon's strange form factors. At
present such calculations remain tremendously challenging because
strange quark effects must inherently originate from disconnected
insertions where the $\overline{s}s$ loops originate from the
QCD vacuum. At present such a calculation remains
prohibitively time consuming. The calculations to date are in quenched
QCD and computed with a large pion mass so some form of extrapolation
to a more physically meaningful situation is required.  Dong and
collaborators~\cite{Don98} carried out a lattice simulation with a
relatively small number of gauge configurations, finding
$\mu_s=-0.36\pm 0.20$ and a positive slope for $G_E^s$ at low momentum
transfer.  Lewis {\it et al.}~\cite{Lew03} carried out a similar
calculation but with a significantly larger number of gauge
configurations and use of a chiral extrapolation to small pion masses,
and found both form factors to be small over a range of momentum
transfer.  Their results were found to be consistent with the existing
data and with a lattice-inspired calculation of Leinweber~{\it et
al.}~\cite{Lei00}.

A variety of other calculations of strange quark form factors have
been carried out with different nucleon models; we will not attempt to
describe them all here.  These include chiral
quark~\cite{Han00a,Han00b,Lyu02} and
soliton-based~\cite{Kim97,Wei95,Sil02} models, in which baryons are
treated as a bound state of constituent quarks surrounded by a meson
cloud, as well as Skyrme-type~\cite{Par91,Par92} models. These models
can eventually provide some guidance as to the origins of potential
strange quark effects when the number and quality of the data points
are sufficient to constrain them.

\subsection{Weak Axial Form Factor and Quasielastic scattering from Deuterium}
\label{sec:axial}

As described in Section \ref{sec:formfactors}, reliable extraction of
the strange quark form factors from the parity-violating asymmetry
also requires a determination of the proton's effective axial form
factor $G_A^e$, which is modified by electroweak radiative corrections
that cannot yet be computed with high precision.  The parity-violating
asymmetry in quasielastic scattering is relatively insensitive to the
strange quark form factors but is predominantly sensitive to $G_A^e$
and can thus provide experimental confirmation of the computed
radiative corrections.

In the simplest, ``static'', approximation where the two nucleons
in the deuteron are treated as stationary, noninteracting
particles, the parity-violating asymmetry in quasielastic
$d(e,e^\prime)$ is
\begin{equation}
A_d = \frac{\sigma_p A_p + \sigma_n A_n}{\sigma_p + \sigma_n} \, ,
\end{equation}
where $A_p$ is defined by equation~\ref{eq:pvee}, and $A_n$, the PV
asymmetry for scattering from the neutron, is written by exchanging
the neutron and proton indices in the form factors.  At backward
scattering angles where the asymmetry is predominantly sensitive to
the magnetic and the axial vector contributions to $A$, it can be
shown that the (isoscalar) strange quark contributions are multiplied
by the isoscalar combination
$G_M^p+G_M^n$ and are thus suppressed, whereas the non-strange (predominantly
isovector) axial vector piece contains $G_M^p-G_M^n$ and is thus not suppressed.
As a result, $A_d$ can be used as a control measurement for $G_A^e$ and its
uncertain radiative corrections.

\begin{figure}
\begin{center}
\includegraphics[width=10 cm]{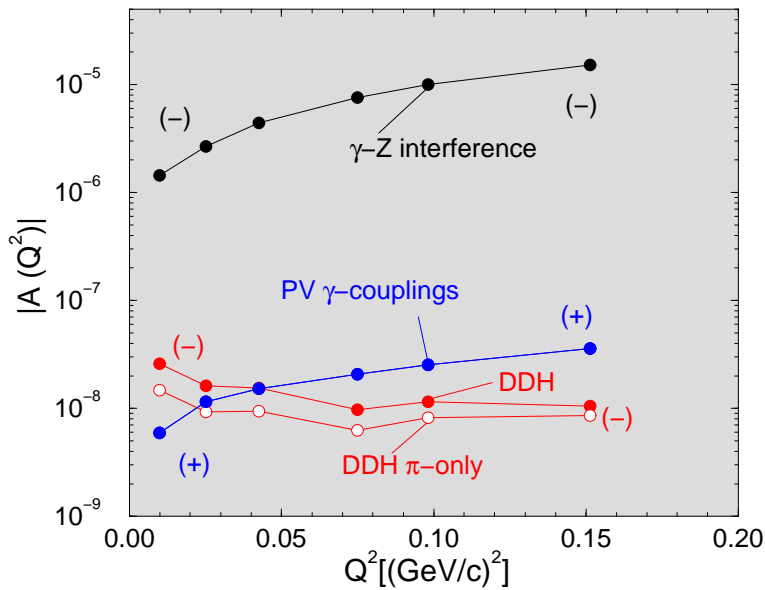}
\caption{Calculation of the parity-violating asymmetry in deuteron
electrodisintegration, from~\protect{\cite{Sch03}}. The process
labeled ``$\gamma$-$Z$ interference'' is the dominant one-body
contribution, the two labeled ``DDH'' are the contributions from the
parity-violating NN interaction, and that labeled ``PV
$\gamma$-couplings'' are the one-body anapole term plus two-body pion
currents. The open (closed) symbols represent positive (negative)
contributions to the asymmetry.}
\label{fig:schiavilla}
\end{center}
\end{figure}

More generally, with moving and interacting nucleons, the asymmetry is
dependent on both the momentum and energy transfer and is modified by
NN interaction effects.  The terms containing the nucleon form
factors can instead be written as a ratio of response functions
\begin{equation}
A_d = \frac{G_FQ^2}{4\pi\alpha\sqrt{2}}\, \frac{W^{PV}}{W^{EM}}
\end{equation}
where
\begin{eqnarray}
W^{EM} &=& v_L R^L(q,\omega) + v_T R^T(q,\omega ) \nonumber \\
W^{PV} &=& v_L R_{AV}^L(q,\omega) + v_T R_{AV}^T(q,\omega) +
v_{T^\prime} R_{VA}^{T^\prime}(q,\omega)
\end{eqnarray}
and the electron kinematic factors $v_T$, $v_L$ and $v_{T^\prime}$
are
\begin{equation}
v_L = \left[\frac{Q^2}{q^2}\right]^2 \, , \> v_T =
\frac{1}{2}\left|\frac{Q^2}{q^2}\right| +
\tan^2\frac{\theta}{2}\, , \> {\mathrm{and}} \> v_{T^\prime} =
\tan \frac{\theta}{2}\left[\left|\frac{Q^2}{q^2}\right| + \tan^2
\frac{\theta}{2}\right]^{1/2} \, .
\end{equation}
The response functions capture both the single nucleon currents and
now also the corresponding two-nucleon currents. Studies of the effect
of the NN interaction on the extraction of these single-nucleon
quantities have been carried out by several authors, generally all
leading to the conclusion that they tend to be small, although this
conclusion depends somewhat on the details of the particular
measurement. In~\cite{Had92}, Hadjimichael, Poulis, and Donnelly
looked at the sensitivity of the asymmetry to the choice of NN
interaction, for example.  They used several models and surveyed a
range of kinematics, finding that at backward angles and moderate
momentum transfers the PV asymmetry varies little with choice of
model, whereas at either low momentum transfer or forward scattering
angles bigger variations between models were found.  The best
sensitivity to the nucleon's axial form factor is at backward angles,
and near the kinematics the first SAMPLE experiment the NN model
dependence appears to be at most a few percent.  More recently,
Diaconescu~{\it et al.}~\cite{Dia01} computed the effect of the
parity-conserving components of two-body contributions on the
parity-violating $e$-$d$ asymmetry using the Argonne V18 potential at
the specific kinematics of the SAMPLE experiment, finding that these
two-nucleon currents modify the asymmetry by up to 3\% in the tails of
the quasielastic distribution and closer to 0.5\% near the
quasielastic peak. Parity-violating two-nucleon contributions to the
hadronic axial response function $R_{VA}^{T^\prime}$ were computed
in~\cite{Hwa81}, and recently again in~\cite{Liu03,Sch03}.  An example
of the magnitude of the parity-violating NN contribution (labeled
``DDH'') relative to the main one-body $\gamma$-$Z$ interference
contribution, taken from~\cite{Sch03}, is shown in
Figure~\ref{fig:schiavilla}.  At the kinematics of interest here,
these effects are about two orders of magnitude below the $\gamma$-$Z$
interference contribution, and are thus negligible.  This is in
contrast to the situation for threshold disintegration or deuteron
radiative capture, where the hadronic PV contributions dominate,
allowing the possibility of measuring the longest range part of the PV
NN coupling, $h_\pi$, via the process $n+p\longrightarrow
d\gamma$~\cite{Bow03}.

In summary, while theoretical determination of $G_A^e$ remains
somewhat uncertain because of hadronic effects at the quark level,
parity-violating electron scattering from deuterium appears to be a
clean probe of $G_A^e$, free from any additional uncertainties arising
from the nuclear environment.

\subsection{Related Observables: Spin, Mass, and Momentum}
\label{sec:otherobservables}

Strange quarks contributions to other observables, such as the
nucleon's spin, mass, and internal momentum distributions are well
documented.  The best source of information on strange quark
unpolarized parton distribution functions, which are a measure of the
fractional internal momentum distributions of the components of the
quark-antiquark sea, is neutrino-induced dimuon
production~\cite{Gon01}.  Neutrinos scattering from either an $s$ or
$d$ quark in a nucleon produce a charm quark and $\mu^-$, and the
charm subsequently decays producing a $\mu^+$.  Tagging the events by
simultaneous detection of the $\mu^+$ $\mu^-$ pairs provides a clean
signature for the charm production. The probability of the charm
production resulting from scattering from an $s$-quark is high, and
the ($x$ or $\xi$) momentum distribution of the cross section can be
directly related to the momentum distribution of the
$s$-quark. Similar arguments apply for $\overline\nu$-$N$ scattering,
where the $\overline{s}$ distribution is determined. The two
distributions have generally been assumed to be the same and global
data fits result in an integrated contribution of approximately 20\%
of the non-strange sea quark distributions or 2\% of the total proton
momentum~\cite{Lai97}.  More recently, the possibility of
non-identical $s$ and $\overline{s}$ distributions is being
investigated as a source of discrepancy between the value of
$\sin^2\theta_W$ reported by the NuTeV collaboration~\cite{Zel02} and
the expectation from global fits to electroweak data~\cite{FIT01}.
While this issue remains to be resolved and new analyses to extract
$s(x)$ and $\overline{s}(x)$ independently are underway, the
integrated combination $s+ \overline{s}$ appears to be relatively
stable~\cite{Gon01}.

Another observable for which there is some indication of strange quark
contributions is the proton's mass, coming from a comparison of the
isospin even $\pi$-nucleon scattering amplitude to the theoretical
prediction for the quantity
\begin{equation}
\sigma \equiv \frac{1}{2M_p}\langle p\vert\hat m\left({\overline u}u + {\overline d}d\right)
\vert p \rangle\
\end{equation}
where ${\hat m}=\frac{1}{2}\left(m_u + m_d\right)$. The experimentally
determined scattering amplitude is measured as a function of momentum
transfer $q$ and extrapolated to $q^2=2m_\pi^2$, the Cheng-Dashen
point~\cite{Che71}, and is notated $\Sigma_{\pi N}$. In the absence of
strange quark contributions, these two quantities should be
equal. Extraction of the experimental value $\Sigma_{\pi N}$ has been
plagued by discrepancies in the data and uncertainties in the
extrapolation technique, and determination of $\sigma$ from theory is
also problematic.  A recent reanalysis using only scattering data
relatively low threshold was carried out by~\cite{Ols00}, resulting in a
value of $\Sigma_{\pi N} = 71\pm 9$~MeV, somewhat higher than the
value of 64$\pm$8~MeV previously determined in~\cite{Sai97}, but lower
than that determined by an update~\cite{Pav99} to the analysis
in~\cite{Sai97}, of 90$\pm$8~MeV. Early determinations of the
theoretical value led to $\sigma = 45\pm 8$~MeV~\cite{Gas91}, but a
more recent determination from lattice QCD results in $\sigma\sim
65$~MeV~\cite{Wri01}, not far from the most recent experimental
determination.  This would indicate no strange quark contribution to
the proton's mass. The variations over time of both the experimental
$\Sigma_{\pi n}$ and the theoretical $\sigma$ demonstrate the
difficulty of extracting a definitive determination of this quantity.

In the last decade a considerable body of data has accumulated on
polarized deep-inelastic lepton-nucleon scattering (DIS), resulting in
a well-determined measurement of polarized structure functions (for a
recent review see~\cite{Fil01}) from which one can deduce the
contributions of sea quarks to the spin of both the proton and
neutron. The proton's spin can be formally written as
\begin{equation}
\frac{1}{2} = \frac{1}{2}\Delta\Sigma + L_q + J_g \, ,
\end{equation}
where the three terms are contributions from quark spin, quark orbital
angular momentum, and gluon total angular momentum, respectively. The
collective body of inclusive polarized lepton scattering data indicate
that only a small fraction, approximately 20\%, of the nucleon's spin
comes from the intrinsic spin of the quarks, the rest arising from
$L_q$ and $J_g$. This result requires integrating the nucleon's spin
structure function over the full range of the quark momentum
distribution, which is not completely measured.  Extrapolations into
the unmeasured region are thought to be under control.  Assuming that
SU(3) flavor is a good symmetry, the quark spin piece $\Delta\Sigma$
can be subsequently broken down into its contributions from $u$, $d$
and $s$ quarks, $\Delta\Sigma = \Delta u + \Delta d + \Delta s$, if
the inclusive scattering data are combined with low energy nucleon and
hyperon beta decay parameters, yielding the result that $\Delta s \sim
-0.1$. A determination that does not require the assumption of SU(3)
flavor symmetry can be found from semi-inclusive kaon production in
spin-dependent DIS, and a result, $\Delta s = 0.03 \pm 0.03{\mathrm
(stat.)}  \pm 0.01{\mathrm (sys.)}$, using this technique has recently
been reported by the HERMES collaboration~\cite{Air04}, although
integrated over only part of the quark momentum distribution.

The most direct measure of $\Delta s$ can in principle be made via
elastic neutrino-proton scattering, a close analog of the
parity-violation technique for extracting the vector form factors. The
nucleon's axial form factor as measured by neutrino scattering is free
of the multi-quark radiative corrections that are present in electron
scattering since they are coupled to a photon exchange between the
lepton and hadron. Precise determination of the absolute cross section
for elastic $\nu$-$p$ scattering is extraordinarily challenging due to
the low cross sections and the difficulty in generating high-quality,
high flux neutrino beams. An analysis of existing data~\cite{Ahr87}
was carried out in~\cite{Gar93}.  The neutrino beam covered a broad
range of momentum transfer centered about an incident neutrino energy
of 1.25~GeV, so an assumption about the $Q^2$-dependence of the axial
strange form factor was made. In addition, a strong correlation
between the strange axial and strange vector form factors was seen.
Recently, Pate~\cite{Pat04} used the neutrino data again to
demonstrate that $\Delta s$ can be better determined in a more global
analysis that combines the neutrino data with the parity violation
data, particularly with that expected to come from the G$^0$ experiment
which covers a similar range of momentum transfer (see
section~\ref{sec:G0} below). The combined data will allow a first
determination of the $Q^2$ dependence $G_A^s(Q^2)$ ($\Delta s =
G_A^s(0)$) over the range 0.45 - 0.95~(GeV/c)$^2$.

\section{The SAMPLE Experiment}
\label{sec:sample-exp}

The main focus of this review is the program of experiments carried
out by the {\sc SAMPLE} collaboration, for which the primary goal has
been to determine the contribution of strange quarks to the nucleon's
magnetic moment. The experiment was thus designed to detect scattered
electrons in the backward direction at low momentum transfer, where
contributions from $G_E^s$ are suppressed both because of kinematics
and because the proton has no net strangeness.  At the {\sc SAMPLE}
kinematics the axial term contributes approximately 20\% to the
asymmetry, and, while the strange quark piece of the axial term
contributes negligibly, the uncertainties in the electroweak radiative
corrections suggested that $G_A^e$ should be determined
experimentally, requiring at least one additional measurement beyond
elastic $e$-$p$ scattering.  As discussed above, quasielastic
scattering from deuterium has similar sensitivity to $G_A^e$, or at
least to the dominant isovector component, but significantly lower
sensitivity to $G_M^s$, and thus provided the second degree of freedom
to independently determine $G_M^s$ and $G_A^e$.

The $e$-$p$ scattering measurements were carried out in 1998, followed
by the first $e$-$d$ scattering experiment in 1999. The first analysis
of the combined data sets indicated that, while strange quarks make up
a small fraction of the proton's magnetic moment, the measured
isovector axial form factor ${G_A^e}^{(T=1)}$ was found to be in
disagreement with the theoretical expectation. As a result, a second
$e$-$d$ experiment was carried out in 2000/2001, at lower beam energy.
The different kinematic conditions provided quite different
experimental systematic uncertainties but similar sensitivity to
${G_A^e}^{(T=1)}$. A recently updated analysis of all three data sets
has now brought both deuterium experiments into good agreement with
theory, with little change to the extracted value of $G_M^s$
~\cite{Ito03,Spa03}.  In what follows we provide a summary of the
experimental measurements along with the most recent updates to the
data analysis and the final results for both $G_M^s$ and $G_A^e$ from
the combined set of experiments. We begin, however, with a general
discussion of polarized beam delivery for parity violation
experiments, with the M.I.T.-Bates beam as our representative example.



\subsection{Polarized Electron Beam and Beam Property Control and
Measurement}
\label{sec:beam}

Parity-violating electron scattering experiments require a
polarized electron beam of
high intensity and
quality and the ability to control and accurately measure
the properties of the beam.  These requirements are driven
both by the statistical and systematic error considerations of the
experiments.

The current generation of parity-violation experiments typically measure
asymmetries with a statistical error $\sim$0.1-1~ppm.
Statistical errors of this precision require luminosities
$\sim 4 \times 10^{38}\ \hbox{cm}^{-2}\hbox{s}^{-1}$ for
reasonable ($<$ 1000 hours) running times.
Currently available
high power hydrogen and deuterium targets have typical target lengths
in the 15-40 cm range, which implies the need for 40 - 100 $\mu$A of
40-80\% polarized electron beam.  Modern polarized electron sources have
achieved these requirements.  The most convenient way to quantify the
impact of polarized electron source capabilities on the experiments'
statistical error is in terms of the figure-of-merit, $P^2I$.  Here,
$P$ is the electron beam polarization and $I$ is the beam current.  For
a given running time it is desirable to maximize this figure-of-merit to
achieve the minimum statistical error.

The impact of beam property control and measurement
capabilities on systematic
errors enters primarily in two ways.  First, one must measure the
beam polarization accurately and ensure that it is longitudinal to
the required precision.  Examples of polarimeters and techniques
to accomplish this are discussed in later sections.  Second, one
must control and measure the helicity-correlated properties of the
beam and properly assess their impact on the measurement.
In an ideal parity-violation experiment, no property of the beam
changes when its helicity is reversed;  in reality, many properties
of the beam such as position, angle, and intensity are observed to
change. This can cause a false asymmetry:
\begin{equation}
\label{eq:Afalse} A_{false} = \sum_{i=1}^{N}
\frac{1}{2 Y} \frac{\partial Y}{\partial P_{i}}
{\Delta P_{i}}.
\end{equation}
Here, $Y$ is the detector yield, $P_{i}$ respresents beam properties
including position, angle, intensity, and energy, and
$\Delta P_{i} = P_{i}^{+} - P_{i}^{-}$ is the helicity-correlation
in those beam properties.
The false asymmetries are typically reduced by
construction of a symmetric detector (to minimize
$\partial Y / \partial P_{i}$) and active feedback to reduce
helicity-correlated beam properties (to minimize $\Delta P_{i}$).
Any residual helicity-correlations in the beam properties are
then corrected for by regular measurement of the dependence of the
detector yield on a given beam property (${\partial Y / \partial P_{i}}$)
and then correction of the measured asymmetry: $A_{corr} =
A_{meas} - A_{false}$.  This corrections procedure is used by
all of the current generation of parity violation experiments

The requirements on the beam property measurement and control
devices are set by the corrections procedure.
The beam properties are measured and
averaged continuously and recorded at every helicity reversal.
The beam property
differences $\Delta P_{i}$ typically have a nearly
Gaussian distribution, with a centroid $\overline{\Delta P_{i}}$ and
a standard deviation $\sigma_{\Delta P_{i}}$.  The centroid
represents
the average helicity-correlated beam property difference for that
parameter, and it must be kept small enough so that the corrections
to the asymmetry are typically only a few percent of the measured
asymmetry.  In practice, active feedback on the beam properties
is necessary to achieve this.  The standard deviation
has contributions from the random fluctuations in
that beam property at the helicity-reversal frequency and also the finite
measurement precision of the beam monitoring equipment.
The error on the
determination of the centroid is $\sigma_{\Delta P_{i}}/\sqrt{N}$, where
$N$ is the total number of difference measurements.
Thus, the standard deviation $\sigma_{\Delta P_{i}}$ must be kept small
enough to allow accurate measurements of the centroid in a reasonable
time period, both for
feedback purposes and in determining the error on the corrections
procedure.

Many of the polarized source techniques and beam control and measurement
methods needed for current experiments were incorporated in the
pioneering parity-violating deep inelastic scattering
experiment of Prescott and collaborators~\cite{Pre79} at SLAC in the
late 1970's.  Techniques were further refined and developed
at MIT-Bates~\cite{Sou90} and Mainz~\cite{Hei89} over the next decade.
In the remainder of this section, we focus our discussion on the polarized
beam techniques and the experience of the SAMPLE collaboration at
MIT-Bates in the late 1990's as a representative example.

\subsubsection{MIT-Bates Polarized Electron Source}
\label{sec:polelec}

In this section, we describe some important details of the MIT-Bates
polarized electron source, starting with a discussion of several features
that are common to all polarized electron sources.
The polarized electrons are produced via photoemission from a
specially prepared gallium
aresenide (GaAs) photocathode.  The special preparations include heat cleaning
of the
photocathode and activation of the crystal via deposition of
cesium and an oxidizer (O$_2$ or NF$_3$)
to create a negative electron affinity
surface.  The negative electron affinity surface allows the electrons
optically excited to the conduction band to escape the cathode.
There are two general types of GaAs photocathodes in use: bulk and strained
GaAs.
Due to the band structure of the GaAs crystal,
the photoemitted electrons can be emitted with a theoretical maximum
of 50\% polarization when 100\% circularly polarized
photons are incident on bulk GaAs.  Strained layer GaAsP photocathodes
break an energy level degeneracy in the valence band; in these
photocathodes, emitted electron polarizations of 100\% are theoretically
possible.  In practice, typical polarizations
of 35-40\% are achieved with bulk GaAs crystals, and 70-85\% for strained
layer GaAsP crystals.  Bulk GaAs crystals can have quantum efficiencies (QE)
$>\sim$1\%, while strained GaAsP crystals typically have much a lower QE.
Lasers of various types
are used to provide the incident light flux necessary to achieve the desired
electron beam currents.

\begin{figure}
\begin{center}
\includegraphics[width=13cm]{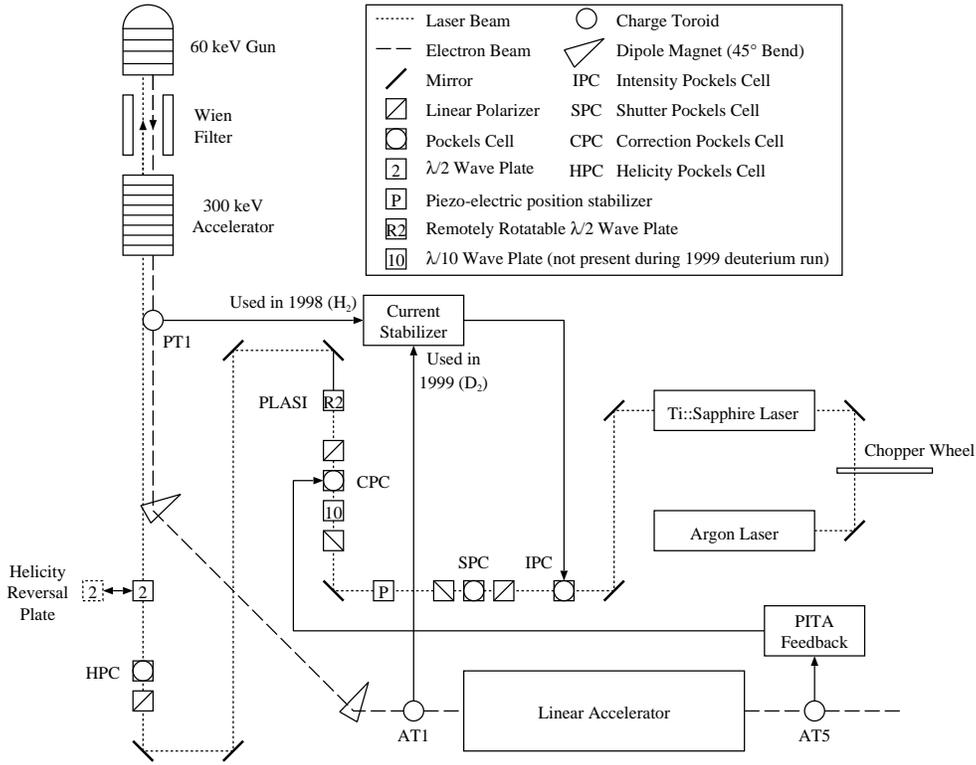}
\caption{A schematic of the MIT-Bates Linear Accelerator polarized
electron source.}
\label{fig:bates_polsource}
\end{center}
\end{figure}

A schematic of the important components of the MIT-Bates polarized electron
source, which was used to provide beams
for the SAMPLE experiment, is shown in Figure~\ref{fig:bates_polsource}.
The MIT-Bates Linear Accelerator is a
pulsed machine, with typically 35 $\mu$s duration electron beam pulses at 600
Hz.  This low duty factor implies a peak current that is about
one to two orders of magnitude higher than that needed for comparable average
beam currents at CW electron accelerators like those at Jefferson Laboratory
or Mainz.  The commercial laser solution adopted at MIT-Bates
consisted of an argon ion laser pumping a Ti:Sapphire laser.  Thermal
lensing effects in the Ti:Sapphire crystal were minimized by chopping
the argon ion laser pump beam with a phase locked electro-mechanical chopper.
The typically available peak powers of about 3.5~W with this system
required that the SAMPLE experiment use bulk GaAs crystals with typical
beam polarizations of $\sim$37\% for production running.  The quantum
efficiency of
the higher polarization
strained GaAsP crystals was too low to achieve the desired
40~$\mu$A average beam current.  Under typical running conditions with
this laser and a bulk GaAs crystal, the MIT-Bates polarized source
peak currents were 4-6~mA, with average currents of $\sim$120~$\mu$A
before chopping of the electron beam in the injector, and $\sim$40~$\mu$A
on the SAMPLE target.  The source produced about 400 C of polarized
electron beam on target over the course of the three SAMPLE production runs.

\begin{figure}
\begin{center}
\includegraphics[width=13cm]{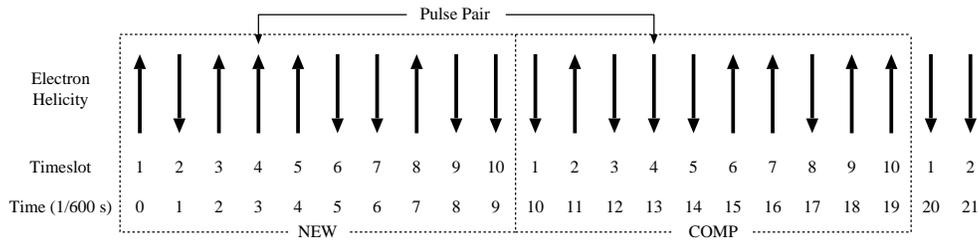}
\caption{Polarization sequence used in the MIT-Bates polarized source.
The first ten states are chosen pseudo-randomly, followed by ten states
that are the complement of the previous ten.}
\label{fig:pol_sequence}
\end{center}
\end{figure}

The time structure of the laser beam was matched to that of the
accelerator with the shutter Pockels cell (SPC) system, consisting
of a Pockels cell between crossed linear polarizers to allow for
electro-optic intensity modulation.  Under normal operating conditions,
this system produced 25 $\mu$s duration laser pulses at 600 Hz.  The
helicity of the laser beam was defined with the helicity Pockels cell (HPC)
system, consisting of a linear polarizer followed by a Pockels cell
set to quarter-wave voltage. Rapid polarization reversal is critical
in parity experiments to reduce sensitivity to slow drifts in
detector and beam properties.  The polarization sequence used in the
MIT-Bates polarized source is shown in Figure~\ref{fig:pol_sequence}.
The electron beam bursts are sorted into ``timeslots'' depending
on their phase in the 60~Hz power line cycle.  Since the accelerator
operating frequency is 600~Hz, there are 10 such timeslots.  For each
of the 10 timeslots, the polarization is selected pseudo-randomly,
referred to as the ``NEW'' sequence.  It is followed by ten states
that are the complement of the previous ten, referred to as the ``COMP''
sequence.  Asymmetries and beam parameter differences
are formed from the ``pulse-pair'' defined
by the NEW and COMP states
for a given timeslot.  Analyzing the data in this way removes
the effect of the dominant 60~Hz power line noise on the experiment.
In addition to reversing the beam helicity with this technique,
an insertable half wave plate, located downstream of the HPC,
was used to manually reverse the beam helicity without changing the
state of the HPC or the helicity control electronics. In the SAMPLE
experiment this was carried out approximately every other day, and
was an important systematic check to assure immunity of the
experiment's detector electronics to the HPC high voltage and related
helicity signals.

A particular aspect of the MIT-Bates polarized source that was important
for SAMPLE running was the intensity stabilizer system.  This was an
active feedback system that reduced the random fluctuations
in the laser intensity.  The system consisted of a Pockels cell electro-optic
intensity modulator (IPC).  The electron beam current was measured at
a point early in the accelerator.  The difference between it and
a reference signal was amplified and applied to the intensity modulator
in order to stabilize the output beam current.  This system suppressed
instabilities in laser intensity, extraction efficiency, and
the first stage of the electron beam transport system.  The system
had a bandwidth of 1~MHz which allowed it to stabilize the beam
current within the 35~$\mu$sec beam pulses. The typical observed
stability of the beam current at the 600~Hz helicity flip frequency
was $\sim$ 0.2 - 0.5\%.

A more complete description of the MIT-Bates polarized electron
source can be found in reference~\cite{Far99}.

\subsubsection{Electron Polarimetry and Spin Transport at MIT-Bates}
\label{sec:beampol}

Two electron beam polarimeters were used during the SAMPLE experiment to
precisely measure the longitudinal component of beam polarization
and for spin transport measurements to manipulate the
spin to a longitudinal state.  The primary apparatus was
a M{\o}ller polarimeter upstream of the SAMPLE detector,
and this was augmented by a transmission polarimeter in an injection chicane early
in the accelerator where the beam energy was 20~MeV.  The
M{\o}ller polarimeter measurements were typically done every 3 days,
while the transmission polarimeter measurements, which were significantly
faster, were carried out daily.

\begin{figure}
\begin{center}
\includegraphics[width=10cm]{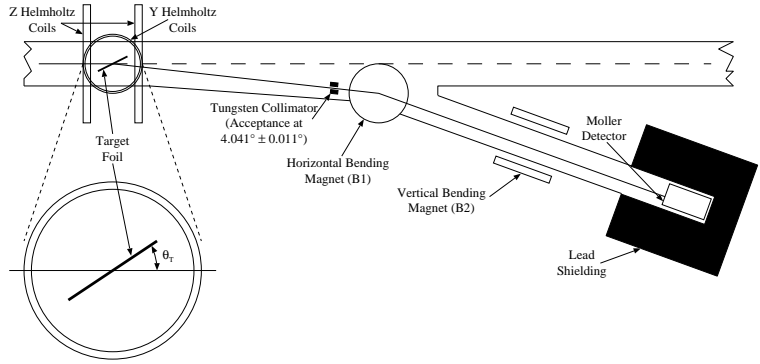}
\caption{Schematic of the SAMPLE M{\o}ller polarimeter.}
\label{fig:SAMPLE-moller}
\end{center}
\end{figure}

The M{\o}ller polarimeter is shown in Figure~\ref{fig:SAMPLE-moller}.
The device was located about 26 m upstream of the SAMPLE target.
The electrons were scattered from a Supermendur foil (49\%
Fe, 49\% Co, 2\% Va), which could
be rotated to any desired target angle.
The target chamber was instrumented with two sets of orthogonal
Helmholtz coil pairs to allow for target polarization both in
the scattering plane and perpendicular to it.
The analyzing power for M{\o}ller scattering is maximum at
$\theta_{CM} = 90^{\circ}$, so the collimator was set to accept
scattered electrons at that angle.  The spectrometer
consisted of two dipoles that gave point-to-point focusing
in both dimensions at a fixed energy of 200 MeV.  The scattered
electrons were detected in a Lucite \v Cerenkov counter.
The resulting signals were integrated over the ($\sim 25-35$~$\mu$s
duration) beam burst.

\begin{figure}
\begin{center}
\includegraphics[width=10cm]{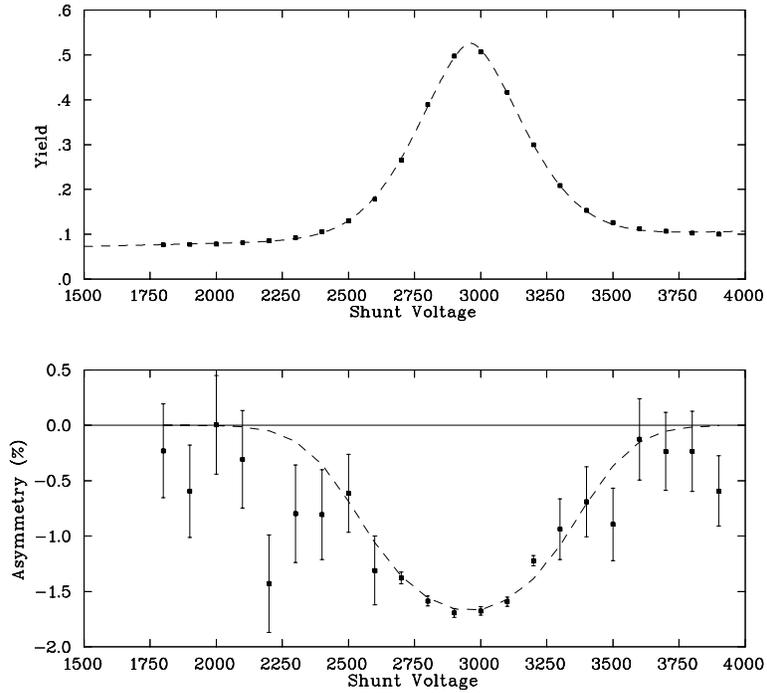}
\caption{A typical momentum scan of the M{\o}ller spectrometer.  The plots
show the yield and asymmetry as a scan through the M{\o}ller peak
is performed.}
\label{fig:mol_run}
\end{center}
\end{figure}

The performance of the M{\o}ller polarimeter satisified the needs
of the SAMPLE experiment.  The M{\o}ller peak was observed by
scanning the spectrometer over a large enough range to include the momentum
of the scattered M{\o}ller electrons and backgrounds from nuclear
scattering and other processes.  Typical observed signal to background
ratios were about 5:1. Yield and asymmetry results from
a typical scan of the spectrometer are shown in Figure~\ref{fig:mol_run}.
The relative statistical error on the
polarization in a 20 minute measurement was about 2\%.  The relative
systematic error from all sources was estimated to be about
4.2\%, with the dominant component of this coming from the 3.2\%
uncertainty in the signal to background extraction.  The
correction for the Levchuk effect~\cite{Lev94} due to scattering
from bound electrons was estimated to be about 2.8\% from a Monte
Carlo simulation of the apparatus.
Typically, a M{\o}ller measurement could take up to 2 hours due
to the need to retune the beam so that it was focused on the
M{\o}ller target, and then
restore regular running conditions afterwards.  It was desirable to have
another polarimetry technique with reduced overhead and which could
be done more frequently since the polarization
tends to vary with time.  Empirically, there is a trend for the polarization
to increase as the GaAs photocathode quantum efficiency decreases.
During the two deuterium runs, more rapid polarization
measurements were carried out with a transmission polarimeter.

\begin{figure}
\begin{center}
\includegraphics[width=13cm]{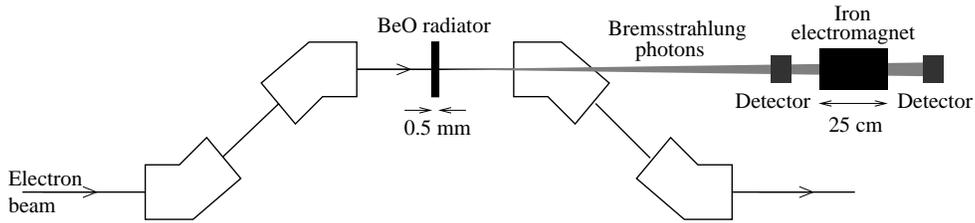}
\caption{A schematic view of the MIT-Bates transmission polarimeter.}
\label{fig:tranpol}
\end{center}
\end{figure}

\begin{figure}
\begin{center}
\includegraphics[angle=90,width=8cm]{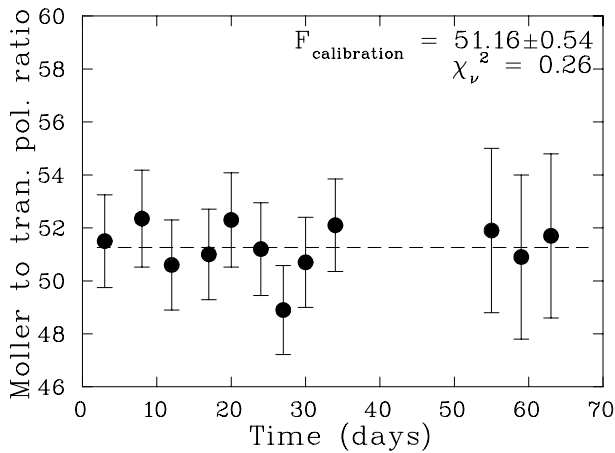}
\caption{The ratio of the polarization from the M{\o}ller polarimeter
and transmission
polarimeter (in arbitrary units) versus day shown
over the course of 60 days.  The cross-calibration was very
stable during the entire period.}
\label{fig:tranpol_results}
\end{center}
\end{figure}

The transmission polarimeter was located in the middle
of the input chicane of the
MIT-Bates accelerator, at a point where the beam energy was 20~MeV.
A schematic of the polarimeter is shown in Figure~\ref{fig:tranpol}.
During normal operations, the beam would pass straight through, but
the beam could be quickly deflected into the chicane for polarization
measurements when desired.  The whole process, including restoring regular
operations, could be completed in 10 minutes.  The transmission
polarimeter consisted of a BeO radiator in the middle of the chicane.
Bremsstrahlung emitted from the radiator was detected in a transmission
style photon Compton polarimeter.  It consisted of two
scintillation counters with an iron electromagnet in between them.  The
longitudinal component of the electron beam spin was transferred
to the bremsstrahlung.  The circular polarization of the bremsstrahlung
was detected via the spin-dependent absorption in the polarized
iron and the resulting asymmetry in the transmitted flux.  The
absolute analyzing power of such a device is difficult to calculate
directly.  Instead, it was determined empirically by cross calibrating
with the SAMPLE M{\o}ller polarimeter.  The
transmission polarimeter functioned very
well as a relative polarization monitor during the run.  It typically
delivered beam polarization measurements with $< 1\%$ relative statistical
error in about 2 minutes of measurement time.  The cross-calibration
relative to the M{\o}ller polarimeter was very stable with respect to
time; results are shown in Figure~\ref{fig:tranpol_results}.

\begin{figure}
\begin{center}
\includegraphics[width=7cm]{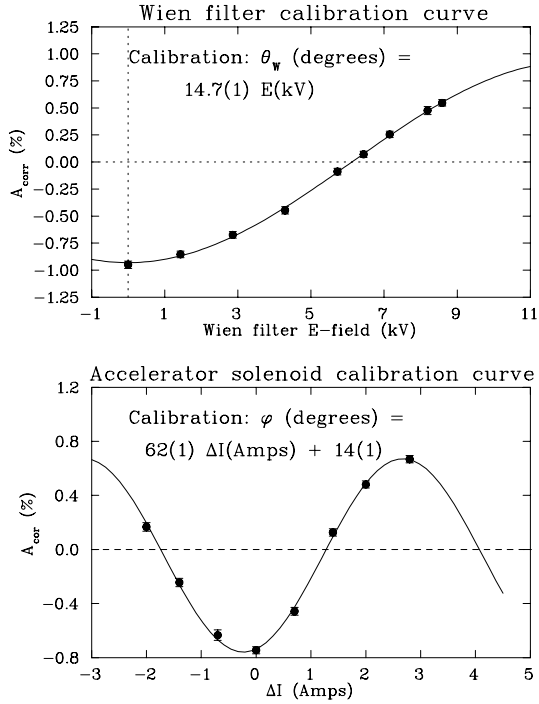}
\caption{In the top figure a calibration curve for the Wien filter is shown
as the M{\o}ller asymmetry as a function of the Wien filter electric field setting.
In the lower figure is a result of the calibration of the
accelerator solenoids resulting from the M{\o}ller asymmetry as a function of the
net deviation of the solenoid magnet currents from their nominal settings.}
\label{fig:spintran_results}
\end{center}
\end{figure}

For the SAMPLE experiment, it was important to ensure that
the transverse beam polarization components were small, because there exists
a parity-conserving left-right analyzing power that can result in a false
asymmetry if the detector geometry is not perfectly symmetric about the
beam line. A careful spin transport procedure was developed for the
experiment, used both to minimize the transverse polarization for regular running
and also to maximize the transverse polarization for dedicated runs
to measure the sensitivity of the experiment to it.  The
SAMPLE beamline was at a 36.5$^{\circ}$
bend relative to the main accelerator, which resulted in a precession
of the electron spin orientation, due to its (``$g-2$'') anomalous magnetic moment,
by 16.7$^{\circ}$ at 200~MeV.  This was compensated for by pre-rotating
the electron spin in the 60~keV region of the accelerator
with a Wien filter (a device with crossed electric and magnetic fields).
However, there also are many focusing solenoid magnets located in the low
energy end of the accelerator which would cause the spin to precess about the
beam axis, and they had to be set to insure that the net precession put
the spin back into the bend plane of the accelerator.  A procedure was
developed to independently calibrate the effect of the Wien filter and
accelerator solenoids on the spin so that they could be used to reproducibly
set the desired spin direction. The procedure~\cite{Pit96} made use
of the SAMPLE M{\o}ller polarimeter, which was
%
%
%
equipped with a rotatable target ladder, allowing the target
to be polarized with components both perpendicular
and parallel to the beam direction in the bend plane.  In terms
of all the relevant angles, the asymmetry between left and
right-handed electrons in the M{\o}ller apparatus can be written as:
\begin{eqnarray}
\label{eq:spintran}
A \;\:&=&\;\:{(P^B P^T / 9)}[{ 1 / (1 + B/S)}] \times \\
&&\;\:[(7\cos\theta_T  \sin  \theta_g  -\sin\theta_T\cos\theta_g)
\sin\theta_W\langle\cos\phi\rangle
-(7\cos\theta_T\cos\theta_g+\sin\theta_T\sin\theta_g
)\cos\theta_W] \nonumber
\end{eqnarray}
Here $P_T\sim 0.080$ and $P_B\sim 0.35$ are the target and beam
polarizations, respectively,  and $S/B$ is the signal to
background ratio of the detected scattered electron yield.
The angles are the Wien precession angle $\theta_W$, the
out-of-plane precession angle $\phi$, the angle between the plane of the
target foil and the beam direction $\theta_T$, and the $g-2$ spin precession
angle $\theta_g$ (16.7$^{\circ}$ for our beamline at 200 MeV).
At a specific target angle ($ \tan \theta_T = 7 \tan \theta_g$),
the first term involving $\phi$ in equation~\ref{eq:spintran} is eliminated.
The second term involves only the Wien angle $\theta_W$, so
the Wien filter can be calibrated directly, independent
of the settings of the accelerator solenoids.  Results of such a
calibration are shown in Figure~\ref{fig:spintran_results}.
The Wien filter calibration is then used to precisely set $\theta_W = 90^{\circ}$, which
eliminates the second term in equation~\ref{eq:spintran}.  The target is
then set at an angle ($\theta_T = -25.6^{\circ}$ for our $\theta_g$)
which maximizes the amplitude of the first term.  Then the dependence
of the angle $\phi$ on solenoid current can be determined, resulting in
a solenoid calibration as shown in Figure~\ref{fig:spintran_results}.
The calibrations obtained from this procedure could be used to
set the spin angles with a precision of
 $\delta \phi = \pm 5^{\circ}$ and $\delta \theta_W
= \pm 1^{\circ}$, allowing the overall spin direction
to be longitudinal to within $2^{\circ}$.  In the SAMPLE experiment this
level of alignment was sufficient to ensure
that any contribution for the parity-conserving left-right analyzing
power is negligible in the SAMPLE experiment.

\subsubsection{Beam Property Measurement and Control}
\label{sec:beam-meas}

\begin{figure}
\begin{center}
\includegraphics[angle=90,width=10cm]{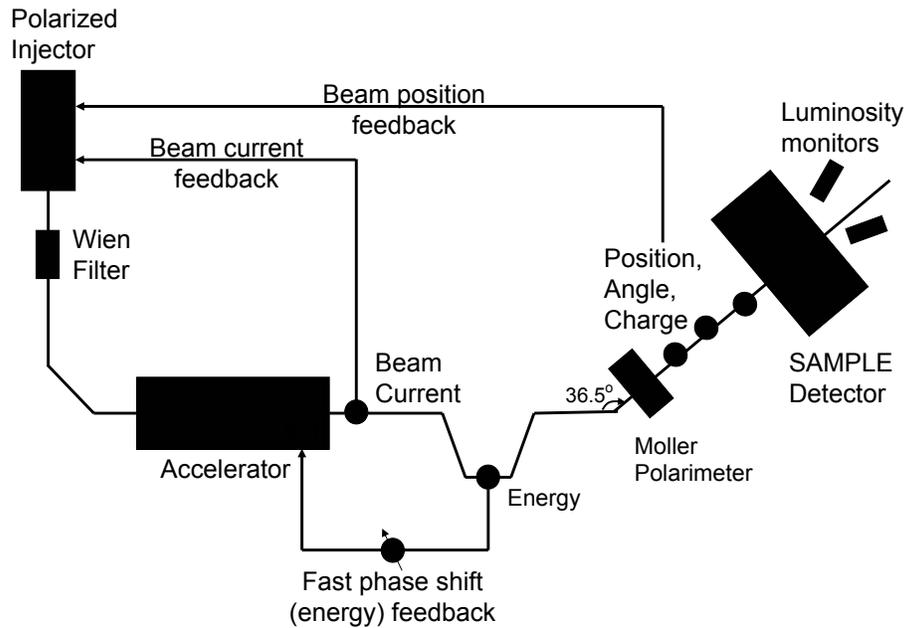}
\caption{Schematic representation of the critical elements for
beam measurement and control in the SAMPLE experiment. The beam current and
beam position feedback systems reduce the helicity correlated charge asymmetry
at the end of the accelerator and the helicity-correlated beam position
differences upstream of the SAMPLE detector.}
\label{fig:bates-layout}
\end{center}
\end{figure}

As outlined in Section~\ref{sec:beam}, measurement and control
of beam properties is a critical part of parity violation experiments.
In this section, we describe how the beam properties for the SAMPLE
experiment were measured and what feedback systems were implemented in order
to control the beam properties at the desired level.
Figure~\ref{fig:bates-layout} shows the critical elements that were used for
beam measurement and control in the SAMPLE experiment.
The beam properties of interest are the beam energy, position and angle
at the SAMPLE target, and beam intensity.  Feedback systems to
reduce the 60~Hz power line noise in the beam energy and helicity-correlations
in the beam position and intensity were implemented.

\begin{figure}
\begin{center}
\includegraphics[width=10cm]{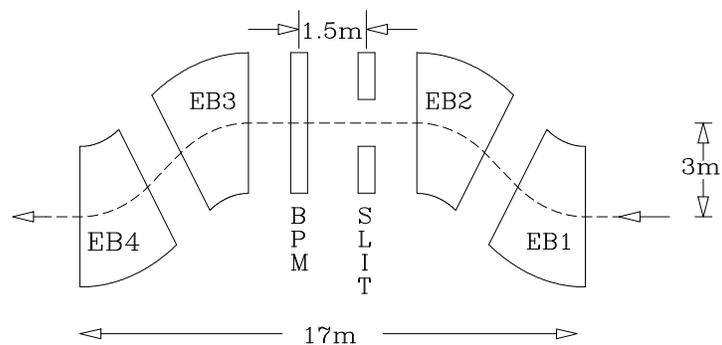}
\caption{Schematic diagram of the MIT-Bates energy chicane showing
the BPM, the energy limiting slits, and the four dipole magnets
that define the magnetic chicane.}
\label{fig:chicane}
\end{center}
\end{figure}

The beam energy was measured at the point of highest dispersion
in the center of a magnetic chicane, a diagram of which is
shown in Figure~\ref{fig:chicane}.
The dispersion at this point was $35\ \hbox{mm}/\%$,
and the beam energy was determined from a measurement of the beam position
at that location.  The dominant fluctuations in the beam energy were correlated
with 60~Hz variations in the electrical power.  This was potentially problematic
for the SAMPLE experiment because when the variations in the energy
were significant enough to cause tails of the beam to scrape on an energy
defining collimator, a significant background in the SAMPLE detector
could be observed.

\begin{figure}
\begin{center}
\includegraphics[width=10cm]{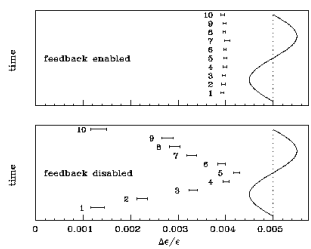}
\caption{Fractional energy change as a function of phase
of the 60~Hz line cycle (``timeslot''). The lower (upper) panel
shows the behavior of the beam with the energy feedback
system disabled(enabled).  The 60~Hz AC line voltage is
superimposed to set the time scale.}
\label{fig:efeed_results}
\end{center}
\end{figure}

To reduce this variation, a feedback system
was implemented~\cite{Bar00}, in which the beam energy for each of the
ten phases (``timeslots'') of the 60~Hz line cycle during the 600~Hz
running was measured independently.   Corrections were then applied
to force the beam energy to be the same for all ten timeslots, thus
greatly reducing the 60~Hz fluctuations.  The energy adjustments were made
by slightly detuning one of the accelerator klystrons away from its
maximum and using a fast phase-shifting device to compensate for the 60~Hz
variations. Results of this system are shown in Figure~\ref{fig:efeed_results}.
The 60~Hz variation in the beam energy was reduced by an order of magnitude, and
this system was also very effective in suppressing slow ($\sim$ seconds or longer)
beam energy drifts of thermal origin.

The beam position and angle at the SAMPLE target were measured
with a pair of microwave cavity position monitors,
similar to those employed at SLAC~\cite{Far76}.  These monitors give a
signal that is proportional to the product of beam position
and beam charge.  The beam charge was measured using a nearby
toroid monitor, the position then resulting as the ratio of the
two signals.  Two (XYQ) monitors
were located 4 and 8 meters upstream of the target so that both
position and angle could be measured in the horizontal and vertical directons.
The monitor signals were integrated and digitized every 25 $\mu$s
beam burst, and helicity-correlated ``pulse-pair'' differences
were formed to continuously measure the difference of the
average beam position for the left and right-handed helicity
states of the electron beam.  Typical standard deviations for these
distributions were $\sim$20-60 $\mu$m, where the dominant contribution
came from the random fluctuations in the beam position at the reversal
frequency.  The intrinsic resolution of the monitors
was small compared to this.

Under normal conditions, the helicity-correlations in the beam position
were large enough that they would lead to large false asymmetries
in the experiment.  These correlations had their origin in the optical elements
in the laser beam path, as determined from extensive study.
The laser beam position was measured with a photodiode segmented into four quadrants.
There were two general classes of effect: direct effects from the
helicity-defining Pockels cell  and secondary effects due to spatial
polarization gradients in the laser light circular polarization created
by the Pockels cell.  The direct effects were observed as helicity-correlated
laser beam angular steering and position shifts when the high voltage
on the Pockels cell was reversed.  Secondary effects were due to spatial
gradients in the circular polarization quality across the laser spot.  When
the laser beam interacted with optically analyzing surfaces (like mirrors)
this would cause a spatially dependent light transmission which appeared
as a position shift of the laser beam.  The secondary effects
were minimized by making the HPC the last optical element in the system,
right before the vacuum window into the electron gun.  The direct effects
were minimized by careful alignment of the Pockels cell, but the residual effects
were still large enough that a dedicated feedback system to minimize
helicity-correlated position differences was necessary.

\begin{figure}
\begin{center}
\includegraphics[width=7cm]{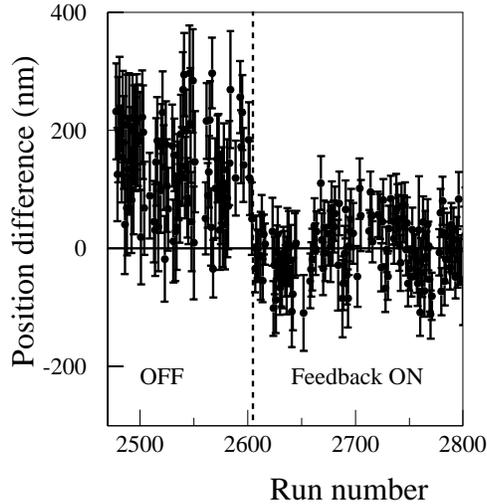}
\caption{Some typical position differences measured in one hour
runs with the helicity-correlated position feedback system on
and off.}
\label{fig:pos_feed_results}
\end{center}
\end{figure}

The position feedback system~\cite{Ave99} was implemented using a plate of
optical glass in a mount with piezoelectric (PZT) steering pads, placed directly
in the laser beam path.
For a small tilt angle, $\theta$, of the glass with respect to the laser
beam there is a pure translation of the laser
beam $\delta = (1 - \frac{n_a}{n_g}) t \theta$, where $n_{a(g)}$ is the
index of refraction of the air (glass) and $t$ is the thickness of the
glass.  The piezoelectric pads could be
driven at the helicity reversal frequency (600~Hz), so it was possible to
use this system to correct for the
helicity-correlations.  In practice, the electron beam position differences were
typically measured over a period of eight hours, resulting in a
precision of about 25~nm.  The PZT system
would then be adjusted to null the
position differences at the SAMPLE target.
Some typical position differences with PZT feedback on and off are shown in
Figure~\ref{fig:pos_feed_results}. Typical position differences with no
feedback were $\sim$50-200~nm, and with
feedback implemented, these position differences were reduced by an order
of magnitude.

\begin{figure}
\begin{center}
\includegraphics[width=7cm]{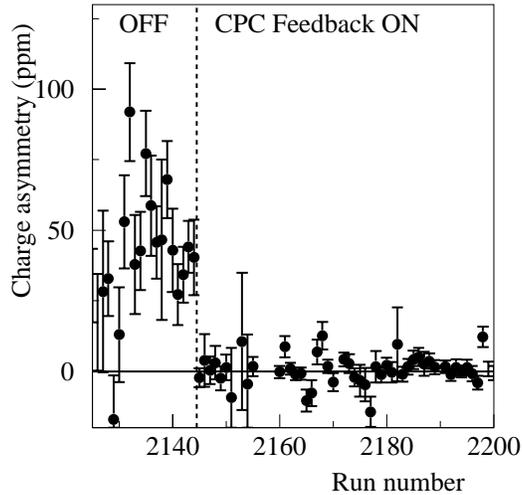}
\caption{Typical intensity asymmetries measured in one
hour runs with and without the CPC feedback system running.}
\label{fig:int_feed_results}
\end{center}
\end{figure}

The net beam charge in each beam pulse was measured non-invasively
with monitors located upstream of the SAMPLE target.
Each monitor consisted of an iron toroid wrapped with wire to detect the
change in magnetic flux as the beam passes through the center.
Helicity-correlations in the beam intensity can affect
the experiment in two ways.  The first is through non-linearities
in the detector phototubes or electronics.  After the integrated detector
signal is normalized to the beam charge, any non-linearity
will show up as a dependence of this normalized yield on beam current.
So, a nonzero beam charge asymmetry would lead to a false asymmetry.
The second, and more significant effect for SAMPLE, resulted from
the beam loading of the MIT-Bates accelerator, or dependence of
beam energy on the beam intensity, which could result in helicity-correlated
energy shifts.
A feedback system (called the CPC, for corrections Pockels cell)
was thus implemented to reduce the helicity-correlated intensity
asymmetry.  The origin of the intensity asymmetry was found to be,
as with the position differences, in the helicity Pockels cell (HPC).
The HPC produces imperfect circularly polarized light resulting in small
residual linearly polarized components in the two states.  These
two states then have different transmission coefficients when they
interact with downstream optical elements, thus resulting in an
intensity asymmetry.  The CPC consisted of linear polarizers
with the same orientation sandwiching a Pockels cell
operating at low voltages.  The CPC could be driven at different voltages
for the two helicity states, so it acted as a helicity-correlated
intensity modulator.  The feedback was carried out by measuring the
intensity asymmetry in a beam charge monitor at the end of the accelerator.
The measurement time was typically about 3 minutes, allowing
the intensity asymmetry to be determined with a precision of $\sim$ 10-20~ppm.
The CPC control voltage was then adjusted to null the intensity
asymmetry.  Typical results with the CPC feedback system on and
off are shown in Figure~\ref{fig:int_feed_results}.  Without feedback
the charge asymmetries were typically 30-70~ppm, but with feedback
they were suppressed by more than an order of magnitude.

The helicity-correlated position and intensity feedback systems
described here were used for all three of the data-taking runs
of the SAMPLE experiment.  Results for the helicity-correlated
beam properties averaged over an entire major run consisting
of about 800 hours are shown in Table~\ref{tab:hel_corr}.  The
resulting values were such that the corrections to the data
for false asymmetries were less than the statistical
uncertainty of the measured parity-violating asymmetries.

\begin{table}
\begin{center}
\begin{tabular}{|c|c|}
\hline
Beam property & Run-average helicity correlation \\
\hline
$\Delta X$ & $7.6 \pm 2.3\ \hbox{nm}$ \\
$\Delta Y$ & $-1.9 \pm 1.4\ \hbox{nm}$ \\
$\Delta \theta_X$ & $-0.6 \pm 0.1\ \hbox{nr}$ \\
$\Delta \theta_Y$ & $-0.9 \pm 0.3\ \hbox{nr}$ \\
$\Delta E$ & $0.14 \pm 0.02\ \hbox{eV}$ \\
$A_I$ & $-0.04 \pm 0.13\ \hbox{ppm}$ \\
\hline
\end{tabular}
\caption{Helicity-correlated beam properties averaged over
about 800 hours for the 1998 SAMPLE LH$_2$ running period.}
\label{tab:hel_corr}
\end{center}
\end{table}


\subsection{The SAMPLE Experimental Setup}
\label{sec:apparatus}

\begin{figure}
\begin{center}
\includegraphics[width=8cm]{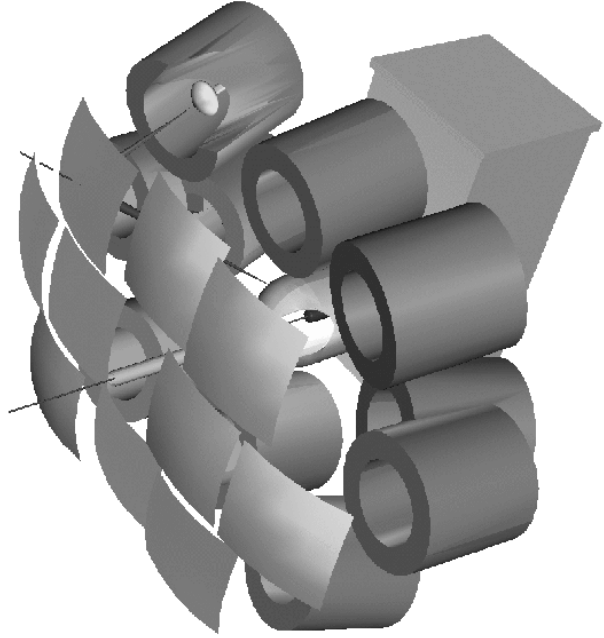}
\caption{Schematic of the {\sc SAMPLE} apparatus. Portions of the scattering chamber and
lead shielding have been cut away for clarity.}
\label{fig:SAMPLE-apparatus}
\end{center}
\end{figure}

A schematic of the {\sc SAMPLE} apparatus is shown in
Figure~\ref{fig:SAMPLE-apparatus}. The polarized electron beam was
incident on a 40~cm long aluminum cell filled with liquid
hydrogen. The scattered electrons exited the target and passed through a
3.1~mm thick hemispherical aluminum scattering chamber lined with
2.5~mm of Pb before entering the volume of air that served as a {\v
C}erenkov medium for the detector.  The detector, the design of which
was based on a similar detector used at the Mainz
Laboratory~\cite{Hei89}, consisted of ten ellipsoidal mirrors that
focused {\v C}erenkov light onto ten 8-inch diameter photomultiplier
tubes. This constituted an azimuthally symmetric detector system with
a solid angle of approximately 1.5~sr, covering scattering angles
between 138$^\circ$ and 160$^\circ$.  The photomultiplier tubes were
encased in Pb cylinders to minimize background from electromagnetic
radiation, and remotely controlled shutters in front of the phototubes
allowed measurement of non-light producing background entering the Pb
cylinders. The detector components and target were encased in a
light-tight box and all metal surfaces were blackened in order to
minimize background from stray light. In the two deuterium
experiments, borated polyethylene shielding was added between the
target and photomultiplier tubes to reduce background from
photo-produced neutrons in the target.

Due to the high count rate in the detector system (more than 10$^8$
s$^{-1}$), the detector signals were integrated over the 25~$\mu$s
long beam pulse of the Bates beam. The energy threshold for {\v
C}erenkov production in air for electrons is 20~MeV, thus the measured
yield in the detector was the integral of all scattered electrons and
positrons above this value.  As a result, background in the detector
had to be measured in dedicated runs. The non-light producing
background was measured regularly throughout the experiments using the
phototube shutters. In the light portion of the signal, an additional
component, approximately 15\%, was due to scintillation light produced
in the air. This component was measured at reduced beam current by
placing a set of auxiliary detectors behind each mirror, requiring a
coincidence between the auxiliary detectors and the phototubes, and
analyzing the pulse-height spectrum of the phototube signals. In the
later experiments an alternative method for determining the
scintillation component at high beam current was developed, which
consisted of measurement of the phototube yields with the mirrors
covered. Both methods are discussed in more detail below.

The cryogenic target system~\cite{Bei96} was designed to rapidly flow liquid
hydrogen or deuterium through the 40~cm long cell in order to minimize potential effects
of beam heating on the target density.
The cryogenic fluid had to be able to absorb the approximately
500 Watts of power deposited by the beam with negligible changes in density on the time
scale of a single asymmetry measurement (16~ms). The target was cooled with a
helium gas refrigerator that delivered 12~K coolant through a counterflow
heat exchanger and was capable of removing
up to 700~Watts of bulk heating. A Chromel ribbon resistive heater immersed in
the target fluid was operated in a feedback loop with the beam current
to maintain a constant heat load on the target that
kept the average temperature of the target constant
to within 0.5~degrees. Measurements to look for fluctuations in
target density were carried out in dedicated periods during each
experiment and they were found to be negligible compared to the statistical
fluctuations of the measured yield.

Small {\v C}erenkov detectors consisting of lucite attached to two inch
photomultiplier tubes were located downstream of the SAMPLE target
at a scattering angle of $\sim 7.5^{\circ}$.  Typically, two to four
of these luminosity monitor detectors were used at various locations about the
azimuth.  Each detector had a factor of three smaller statistical error
than the combination of the ten main SAMPLE detectors, and as a
result they were a sensitive monitor of potential fluctuations in target density.
These monitors were also used as a null asymmetry monitor. The detected
signal in these monitors was primarily from very forward scattered electrons at
low momentum transfer and/or from electromagnetic showers,
for which the expected asymmetry is significantly smaller
than the main parity-violating signal of the experiment.

The raw photomultiplier tube signals were sent through
current-to-voltage amplifiers prior to integration over the 25~$\mu$s
beam pulse. The integrated voltages were dithered with a small
additional random voltage (which was subtracted in analysis) before
being digitized in order to remove the potential effects of
differential nonlinearities in the 16-bit analog-to-digital converter
modules used in the readout system.  The data stream was read out
during the 1.5~ms between beam bursts so the data acquisition system
was free of deadtime. The data stream included not only the ten
photomultiplier tubes but also beam current and position monitors
along the accelerator that would indicate potential helicity
correlated beam properties concurrent with the measurement.  The
helicity of the beam was flipped randomly in a sequence of ten beam
pulses, reversed for the next ten pulses, and then paired to form ten
asymmetry measurements. Approximately one in 39 beam pulses was
blanked to keep track of potential baseline shifts, and the beam
helicity was reported to the data stream after digitization of the
analog signals, and in a symmetric fashion, in order to avoid
electronic cross talk or potential helicity correlated baseline
shifts.  The ten ``time-slot'' asymmetries were averaged together, and
then averaged over a one-hour period.

\subsection{Data Analysis}
\label{sec:data-analysis}
In the first pass through the data analysis, the raw detector asymmetries were
computed, as well as each detector's sensitivity to the measured beam properties using the natural,
helicity-uncorrelated motion of the beam over a one-hour period. In a second pass through the analysis, the
measured asymmetry was corrected for false asymmetries arising from helicity-correlated beam motion using a
linear regression procedure on the detector yields on a pulse-by-pulse basis. The linear regression accounts
for correlations between parameters as well as correlations with the detector yields, and the procedure is
mathematically equivalent to subtraction of a false asymmetry arising from the helicity correlated beam
properties.

\begin{figure}
\begin{center}
\includegraphics[angle=90,width=8cm]{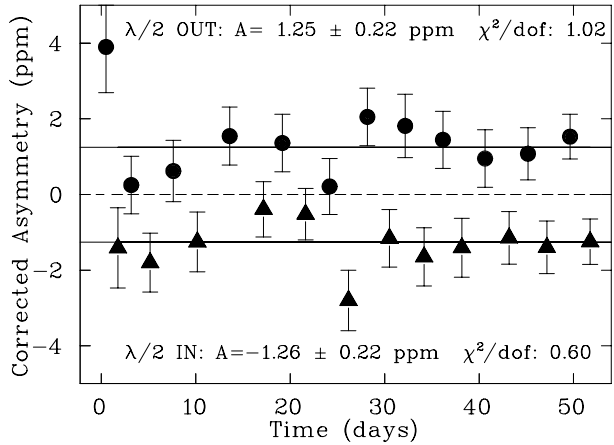}
\caption{Averaged asymmetries for each state of the half wave
plate are shown as a function of time.  The asymmetries in this
plot is the measured asymmetry of the sum of
the 10 photomultiplier tube signals corrected
for helicity-correlated beam properties with the linear regression
procedure described in the text.  The solid lines show the overall
average value of the asymmetry in each of the two states, IN (triangles)
and OUT (circles). It should be noted that this phototube sum signal
is not the quantity used to construct the final experimental asymmetry,
rather the statistically weighted average of the ten
phototube signals was used.}
\label{fig:halfwave_1998}
\end{center}
\end{figure}

With one exception (see below), the corrections made to the asymmetry
were always less than its statistical uncertainty.
The same corrections procedure was also applied to the
luminosity monitors. Due to the small momentum transfer of the accepted events in
these monitors, the expected asymmetry was smaller than the precision
with which it was measured, so these monitors could be used
to assess performance of the corrections procedure in eliminating
false asymmetries and to quantify the errors assigned to the corrections
procedure.  After the corrections were applied to the data,
the detector asymmetry values were sorted based on state of the
insertable half wave plate, which was reversed every few days.
In Figure~\ref{fig:halfwave_1998}, the expected reversal of the
sign of the measured asymmetry is clearly seen in the data.
The net correction, averaged over ten mirrors, applied to the
hydrogen (raw asymmetry) data was 0.07~ppm with an assigned systematic uncertainty
of 0.05~ppm. The systematic uncertainty was estimated by assuming that all
of the residual asymmetry of the luminosity monitors, measured to be
0.05$\pm$0.05~ppm, was due to any error in the corrections procedure.
It should be re-emphasized that the corrections were
computed and applied on a detector-by-detector basis, although the correction
varied little from one detector to the next. The equivalent
quantity in the 200~MeV deuterium run was 0.07~ppm with an assigned
uncertainty of 100\% of the correction.  While in each of these
runs the contribution from beam helicity correlations to both the asymmetry
and its systematic uncertainty are very small, in the third SAMPLE run,
the corrections were larger due to a significant helicity
correlated transmission asymmetry of the beam. This issue is described in
more detail below.

%

Once the measured asymmetry has been corrected for false asymmetries
due to the beam, other dilution factors and potential sources of
yield in the asymmetry that arise from
physics processes other than elastic $e$-$p$ scattering
must be accounted for. The largest dilution factor is that due to the
beam polarization, which was 35-40\% in each of the three
experiments. The remaining corrections can be summarized in the following
expression:
\begin{equation}
\label{eq:Aexp}
A_{exp} = \frac{R_c}{P_B f_l f_c\left(1-f_{\pi}\right)}
\left[A_O - \left(1-f_l\right)A_C - f_{\pi} A_{\pi}\right]
\end{equation}
where $P_B$ is the electron beam polarization, $R_c$ is an electromagnetic
radiative correction due to emission of bremsstrahlung, $f_l$ is the fraction
of the yield due to light incident on the photomultiplier tubes, $f_c$ is the fraction
of the light-producing signal due to {\v C}erenkov radiation, and
$f_{\pi}$ is the fraction of the {\v C}erenkov light due to $\pi$-production
in the target. The quantity $A_O$ is the asymmetry measured under normal
running conditions, after corrections for beam helicity correlations, and
$A_C$ is the asymmetry measured with shutters in front of the photomultiplier
tubes to block incident light. In the first two {\sc SAMPLE} experiments,
approximately one quarter of the data taking was devoted to these ``shutter-closed''
measurements.  In the third, this fraction was reduced to 10\%, but
additional measurements of the {\sc CLOSED} asymmetry were made by enhancing its
contribution
with a piece of plastic scintillator placed in front of each photomultiplier tube.
The light fraction $f_l$ was determined with high precision from these data
and contributed a negligible uncertainty to the dilution corrections.
While in each of the three measurements the average background asymmetry
was consistent with zero, mirror-by-mirror variations resulted in
the need to handle this contribution differently in the three experiments,
as discussed in more detail below.

\begin{figure}
\begin{center}
\begin{minipage}[b]{7.5cm}
   \centering
   \includegraphics[height=5cm]{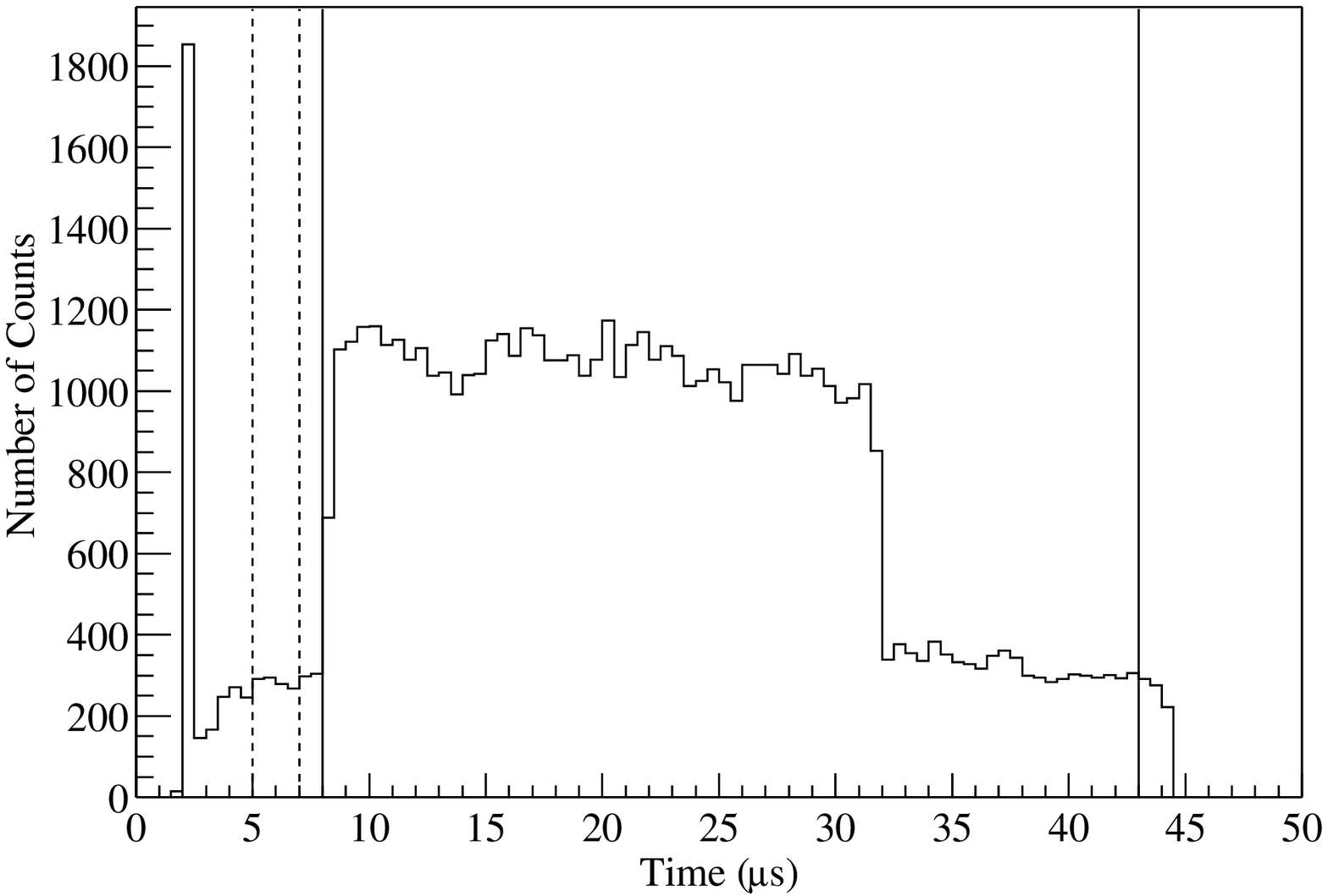}\\
\end{minipage}
\begin{minipage}[b]{7.5cm}
   \centering
   \includegraphics[height=5cm]{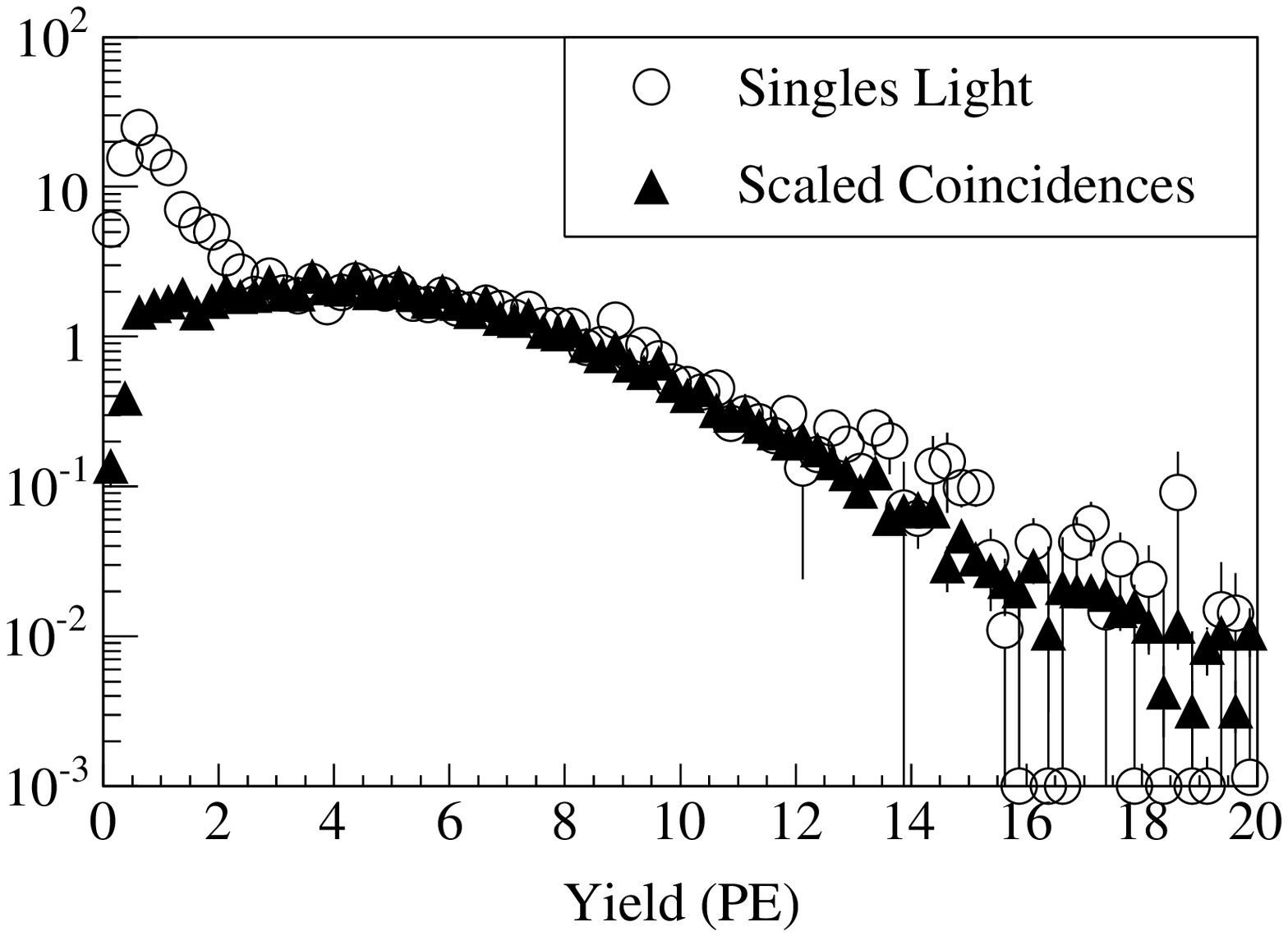}\\
\end{minipage}
\caption{(a) Coincidence event timing spectrum during low-current measurement
of the {\v C}erenkov fraction $f_c$, showing the period of time prior to beam
delivery (dashed lines) used for dark rate determination. (b) Pulse height spectra for
the singles (circles) and coincidence events (diamonds) after correction for dark
current, dead time, and the edge-effects of the mirrors.}
\label{fig:pulsecounting}
\end{center}
\end{figure}

In each of the three {\sc SAMPLE} experiments
the {\v C}erenkov fraction $f_c$ was determined with a set of dedicated measurements
in which the beam current was reduced to a level such that individually scattered
electrons could be counted and the pulse height distributions of the
photomultiplier signals recorded. One in ten beam pulses contained enough
beam charge that statistically meaningful integration measurements could also
be made in order to monitor that beam properties were comparable to those during
normal high-current running (these pulses were used only for monitoring beam
properties and not included in the data stream.)
Identification of the {\v C}erenkov component
of the pulse-height spectrum was made by requiring a coincidence between
a plastic scintillator placed behind each mirror and its corresponding
photomultiplier tube. The {\v C}erenkov fraction was then defined to be
the integrated yield in the coincidence pulse-height
spectrum relative to the yield in the singles spectrum, after normalization
to integrated beam charge and correction for dark current contributions
and dead time effects. An additional correction was made to the coincidence
pulse-height spectrum due to the fact that the scintillators covered only
the central portion of the mirrors. Under normal integration running conditions
and in the singles event pulse height distribution,
the spectrum of {\v C}erenkov photons per incident electron is slightly distorted
because the edges of the mirror do not have full acceptance of the scattered
electron's {\v C}erenkov cone. This distortion of the {\v C}erenkov spectrum
was simulated and resulted in an enhancement of the measured $f_c$ by 5\%.
Figure~\ref{fig:pulsecounting}(a) shows a typical spectrum of coincidence
count rate as a function of time where the 25~$\mu$s beam pulse can be
clearly identified. In Figure~\ref{fig:pulsecounting}(b) the pulse height
spectra for singles and coincidence events are shown for one detector.

In the two deuterium experiments an alternate method of determining $f_c$
was developed that did not require reducing the beam current. Covers
were placed over the mirrors in order to determine what fraction of the
light yield in a given phototube did not involve direct reflection off
of its corresponding mirror. This method was combined with a simulation of
scintillation production in the air between the mirrors and photomultiplier
tubes to determine how much of the scintillation light seen by a particular
phototube reflects off of its own mirror compared with adjacent ones.
This method, while relying on simulation for a small part of the correction,
was found to be in good agreement with the low beam current method and also
could be carried out without changing beam conditions. Further details
regarding the two methods for determining $f_c$ can be found in~\cite{Has03}.
The largest source of systematic uncertainty in the dilution corrections
came from the determination of $f_c$, for which a relative uncertainty of
4\% was assigned. When the low current data were used to determine $f_c$,
this uncertainty was due largely to the statistical uncertainty of the
measurement. The largest source of systematic error was in the determination
of the beam charge for the very low current pulses. With the alternate method
of covering the various detector mirrors, the dominant uncertainty was due to
the assumptions made about the source of the scintillation light, and was also
a few percent.

\begin{figure}
\begin{center}
\includegraphics[width=10cm]{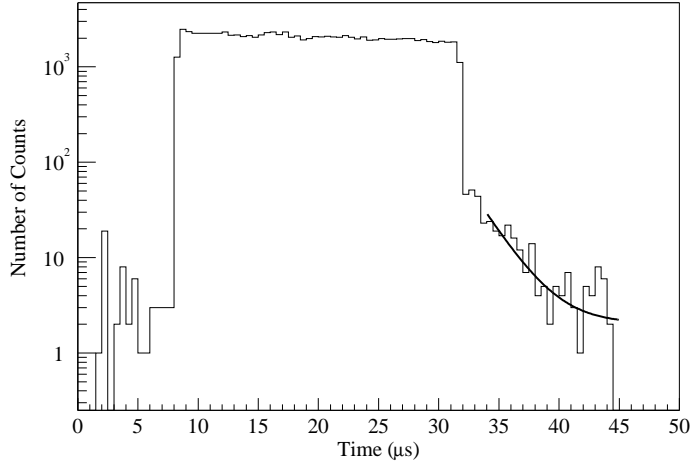}
\caption{Coincidence event timing spectrum during low-current measurement
of the {\v C}erenkov fraction $f_c$, showing the period of time after the
beam pulse delivery used to estimate the rate of events due to muon decay
from $\pi^+$ photoproduction. The solid line indicates a fit to the count rate
of the form $A+B\, exp[-(t-t_0)/\tau_\mu]$.}
\label{fig:pionrate}
\end{center}
\end{figure}

With a beam energy of 200~MeV, a small fraction of the
produced {\v C}erenkov light was due to threshold photo-production of
pions, and the parity-violating asymmetry for such processes has not
previously been measured.  It  has been computed by several
authors~\cite{LiH82,HaD95,Che01a,Che01b}, leading to a typical value
of $\vert A_{\pi}\vert < 2\times 10^{-7}$,
which would be a negligible contribution to the {\sc SAMPLE} measured
asymmetries, so the third term in equation~\ref{eq:Aexp} has been
neglected. However the small dilution correction is still required, and
$f_{\pi}$ was determined within the context of a {\sc GEANT}~\cite{GEANT}
simulation. Photo-produced pions in the target do not contribute directly
to the {\v C}erenkov signal, but only through their decay products. The two
photons from the decay of neutral pions produced in the target can produce
electromagnetic showers that result in a small {\v C}erenkov signal.
Positively charged pions create muons whose decay products can have
sufficient energy to produce {\v C}erenkov light in the detector.
Because of the 2.2~$\mu$sec lifetime of the muon, these events were seen
by the detector system in the dedicated low beam runs, as shown in
Figure~\ref{fig:pionrate}, and the rate
estimated from the data is consistent with that determined through
simulation. In the case of a deuterium target, two additional contributions
are important. First, negatively charged pions are also produced along with
their decay muons, but these muons are likely
radiatively captured in the target before decay. The approximately 100~MeV
photon released in the radiative capture process~\cite{Bae77} can shower
and create a small {\v C}erenkov signal. Finally, the largest source of
rate from pion production in a deuterium target is coherent $\pi^0$
production. This process was recently measured at the Saskatoon
Laboratory~\cite{Ber98} and was found to be significantly enhanced
relative to incoherent production on a single nucleon~\cite{Ber98b},
due to $N$-$N$ rescattering effects~\cite{Bea97}.

The electromagnetic radiative corrections were also computed within the
context of the {\sc GEANT} simulation, using a spin-dependent modification
to~\cite{MoT69}. In addition to the energy loss that results in a reduction
in the four-momentum transferred to the target, electrons emitting
radiation prior to interaction with a target nucleus can undergo a spin-flip,
effectively depolarizing the incident beam. Such depolarization effects
were implemented using the calculations in~\cite{Kuc83} and~\cite{Ols59}
for internal and external bremsstrahlung, respectively. Details of
the computation of the radiative effects can be found in~\cite{Spa01}.

In the {\sc GEANT} simulation, scattered electrons were generated uniformly in
energy, angle and along the length of the 40~cm target. Energy loss due to
straggling in the aluminum entrance window to the target, and in the thickness of
liquid hydrogen upstream of the randomly chosen interaction point,
was accounted for before computation of the scattered electron
kinematics. Each scattered electron was assigned a cross section
and a parity-violating asymmetry, and propagated through the target
exit windows and the scattering chamber. A detection efficiency based on
the velocity of the outgoing electron and the path length of the event's
track in the {\v C}erenkov medium was combined with the computed cross
section as an event weight. Scattered electrons were required to be incident
on a mirror and the generated {\v C}erenkov light was required to have a trajectory
that, when reflected from a mirror, would be incident on a photomultiplier tube.
For the hydrogen target only elastic
$e$-$p$ scattering was evaluated. For the deuterium target, the radiative
effects were evaluated separately for
elastic and inelastic $e$-$d$ scattering,
which were then combined, weighted by their relative cross sections.

Table~\ref{tab:dilutions} contains a
summary of all the dilution corrections for the three experiments, averaged
over the ten detectors. A summary of the systematic uncertainties for each of the
three measurements is shown in Table~\ref{tab:systematics}.

\begin{table}
\begin{center}
\begin{tabular}{|c|c|c|c|c|c|}
\hline
Data set & $f_l$ & $f_c$ & $(1-f_{\pi})$ & $R_c$ & $P_B$\\
\hline
H$_2$ 200~MeV & 0.83 & 0.86 & 0.96 & 1.13 & 36.2\% \\
D$_2$ 200~MeV & 0.71 & 0.89 & 0.90 & 1.12 & 35.7\% \\
D$_2$ 125~MeV & 0.76 & 0.90 & 1.00 & 1.09 & 38.9\% \\
\hline
\end{tabular}
\caption{Summary of dilution corrections for the three {\sc SAMPLE} experiments. The corrections, which varied
a small amount for each of the ten detectors, were applied on a detector-by-detector basis. The light fraction
$f_l$ and Cerenkov fraction $f_c$ were determined directly from experimental data as described above, whereas
$f_{\pi}$ and $R_C$ had to be determined from the GEANT simulation.} \label{tab:dilutions}
\end{center}
\end{table}

\begin{table}
\begin{center}
\begin{tabular}{|c|c|c|c|}
\hline
 Source & & $\delta A/A$ (\%) & \\
\hline
& H$_2$ 200 MeV & D$_2$ 200~MeV & D$_2$ 125~MeV \\
\hline
Beam polarization & 4 & 4 & 3 \\
{\v C}erenkov fraction & 4 & 3 & 3 \\
Radiative correction & 3 & 3 & 3 \\
Pion contamination & 2 & 2 &  \\
\hline
Net dilution factor & 7 & 6 & 5 \\
Background asymmetry subtraction & 13 &  & 5 \\
Asymmetry Corrections Procedure & 6 & 5 & 5 \\
Luminosity Monitor Asymmetry  & 1 & 2 & 14 \\
\hline
Total (quadrature sum) & 16 & 9 & 17 \\
\hline\hline
\end{tabular}
\caption{Summary of systematic uncertainties for the three data sets, indicated
as a relative uncertainty in the measured asymmetry.}
\label{tab:systematics}
\end{center}
\end{table}

\subsection{Results from the Hydrogen Data}
\label{sec:hydrogen}

A preliminary analysis of the hydrogen data was published
in~\cite{Spa00}.  Both the analysis and the theoretical expectation
were since refined, as described above: the main modification to the
theoretical expectation was to recast the one-body weak radiative
corrections using the $\overline{\mathrm{MS}}$ value for $\s2w$, which
resulted in a significant reduction because of their dependence on
$(1-4\s2w)$.

\begin{figure}
\begin{center}
\includegraphics[width=10cm,angle=90]{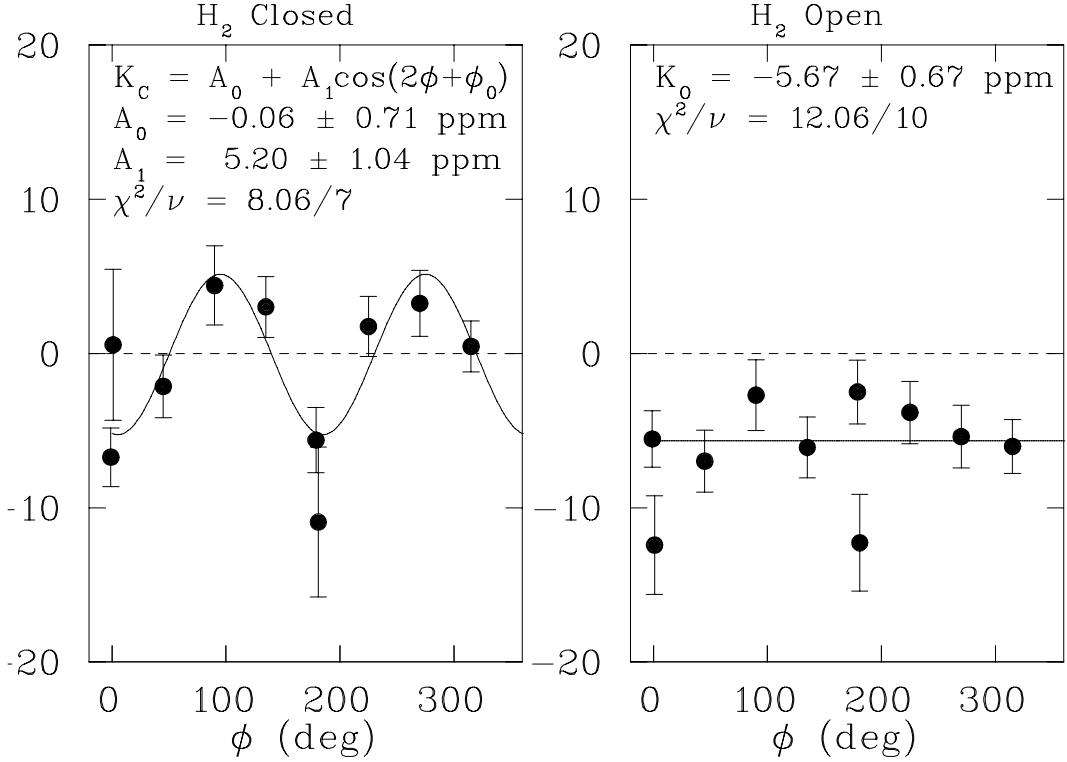}
\caption{Contributions to the asymmetry, corrected for the appropriate dilution
factors, from the CLOSED and OPEN shutter data. The CLOSED contribution was
best fit to a $\cos2\phi$ function while for the OPEN data a weighted average
of the ten detectors was used.}
\label{fig:asym_phi}
\end{center}
\end{figure}

Differences between the first analysis and the subsequent one are
discussed in detail in~\cite{Spa01} and~\cite{Spa03},
primarily coming from use of the {\sc GEANT} simulation to compute the radiative
corrections and pion dilution
correction (as described in section~\ref{sec:data-analysis}),
as well as the treatment of the shutter {\sc CLOSED}
background. The measured background asymmetry, when averaged over the ten detectors, of $-0.62\pm 0.69$~ppm
was consistent with zero but the detector-by-detector distribution was not flat. It had a dependence on the
azimuthal angle $\phi$ about the beam axis, and a fit to a function $f(\phi)=A_0+A_1\cos(2\phi+\phi_0)$ fit
the data with significantly better $\chi^2$/d.o.f (8.0/7) than the presumed flat distribution (33.3/9). This
$\phi$-dependent behavior is not evident in the shutter {\sc OPEN} data with any statistical significance.
Therefore the {\sc OPEN} data were combined in a weighted average of the ten detectors, and the fitted value
of $A_0=-0.06\pm 0.71$~ppm was subtracted from the {\sc OPEN} asymmetry to yield the result
\begin{equation}
A_p(Q^2=0.1) = -5.61\pm 0.67\pm 0.88\, {\mathrm{ppm}}\, ,
\end{equation}
where the first uncertainty is statistical and the second is systematic,
including the uncertainty in the background asymmetry.
The two distributions as a function of azimuthal angle are shown in
Figure~\ref{fig:asym_phi}.

The {\sc GEANT} simulation was also used to compute the theoretical value to which
the data were compared, in order to account for finite acceptance and target
length effects in the detector. Because of the integrating nature of the detector
system, the simulation was not used to compare directly to the asymmetry data.
Dipole form factors were used for the
neutron magnetic and proton electromagnetic form factors in computing the $e$-$p$
cross section, and the Galster parameterization for $G_E^n$. Uncertainties in
the predicted asymmetry due to the electromagnetic form factors are small:
they are dominated by the uncertainties in $G_M^n$, for which
a 3\% uncertainty was assumed, and $G_M^p$, for which the uncertainty
was assumed to be 2\%. The resulting effect on $G_M^s$ is 0.05.
The effect of averaging over
the target length and acceptance was to reduce the predicted value of the asymmetry by 3\%.
Other required quantities for computing the asymmetry were $\sin^2\theta_W = 0.23113$,
$G_F=1.16637\times 10^{-5}$~GeV$^{-2}$~\cite{PDG03}, ${G_A^e}^{(T=0)}=-0.07$.
The resulting theoretical asymmetry is
\begin{equation}
A_p = -5.56 + 3.37 G_M^s + 1.54{G_A^e}^{(T=1)}
\end{equation}
where the asymmetry is in parts per million and the form factor $G_M^s$ is
in nuclear magnetons. Combining the measured hydrogen asymmetry with the
theoretically determined value of ${G_A^e}^{(T=1)}=-0.83\pm 0.26$ results in
\begin{equation}
G_M^s(Q^2=0.1) = 0.37 \pm 0.20 \pm 0.26 \pm 0.07
\end{equation}
where the last uncertainty is due to knowledge of the electromagnetic form
factors and of the electroweak radiative
corrections to ${G_A^e}^{(T=1)}$. The results for individual detector
types are listed in Table~\ref{tab:h200}.

\begin{table}
\begin{center}
\begin{tabular}{|c|c|c|c|c|c|c|}
\hline
Det.~\# & $\langle\theta \rangle$ & $A_0$ & $A_1$ & $A_2$ & rel. weight & $A_{meas}$ \\
\hline
 & (deg) & (ppm) & (ppm) & (ppm) & & (ppm) \\
 \hline
1,3,8,10 & 143 & -5.49 & 3.31 & 1.52 & 0.48 & -5.70$\pm$0.96 \\
2,9 & 152 & -6.11 & 3.73 & 1.69 &  0.19 & -4.14$\pm$1.52\\
4,7 & 136 & -4.90 & 2.93 & 1.37 & 0.24 & -4.14$\pm$1.37\\
5,6 & 158 & -6.53 & 4.03 & 1.80 & 0.09 & -12.29$\pm$2.23\\
\hline
\end{tabular}
\caption{Summary of experimental results and theoretical expectation for the asymmetry,
in ppm, broken down by detector type for the SAMPLE 200~MeV hydrogen data.
The quantities $A_0$, $A_1$ and $A_2$ are the leading term and the coefficients
multiplying $G_M^s$ and ${G_A^e}^{(T=1)}$, respectively. The relative weights
were determined by the uncertainty in the measured asymmetries for each detector
type. The uncertainty in the experimental result is statistical only: the uncertainty
in the background correction was treated as a common systematic error.}
\label{tab:h200}
\end{center}
\end{table}

\subsection{Results from the Deuterium Data}
\label{sec:deuterium}

Analysis of the 200~MeV deuterium data, originally presented
in~\cite{Has00}, has also since been updated~\cite{Ito03}.
When combined with the result in~\cite{Spa00}, the data indicated a
value of $G_M^s$ consistent with zero, but a radiative correction
to ${G_A^e}^{(T=1)}$ that was significantly larger than anticipated from
theory. A new analysis of both the theoretical expectation
and the experimental data were carried out. Three main improvements
have led to a new result for ${G_A^e}^{(T=1)}$ that is now in good agreement
with~\cite{Zhu00}.
As described above and in~\cite{Ito03}, improved determination
of the electromagnetic radiative corrections and {\v C}erenkov
fractions, as well as accounting for coherent $\pi^0$ production,
resulted in an increase of the dilution correction by 11\%. The measured
background asymmetry was consistent with zero and did not exhibit the
$\phi$-dependent behavior seen in the hydrogen data. As a result no
correction for background asymmetry was applied, resulting in a
dilution-corrected asymmetry of
\begin{equation}
A_d = -7.77 \pm 0.73 \pm 0.72\, {\mathrm{ppm}} \, .
\end{equation}

In the determination of the theoretical asymmetry to which the result should
be compared, again the {\sc GEANT} simulation was used to identify finite acceptance
effects. For the theoretical model, parity-violating and parity-conserving
response functions provided by Schiavilla were used that include both quasielastic
scattering and threshold breakup of deuterium. The latter process contributes
approximately 5\% to the experimental yield, with an asymmetry comparable in
magnitude to quasielastic scattering. The calculation, based on that
described in~\cite{Car02} and~\cite{Sch03}, used the
Desplanques, Donoghue and Holstein~\cite{DDH80} parameterization of parity-violating
meson exchange coupled with the Argonne {\sc V18} $N$-$N$ potential~\cite{Wir95}.
The anapole contributions were based on~\cite{Hax89}.

In the computation of the theoretical asymmetry it was also necessary to include
a contribution from elastic $e$-$d$ scattering, which is only 1-2\%  of the
detector yield but for which the asymmetry expected to be comparable in
magnitude but opposite in sign. It is expected that elastic $e$-$d$ scattering has
considerable sensitivity to strange quarks~\cite{Pol90}. The elastic asymmetry was
evaluated using a global fit to deuteron form factor data~\cite{Bal00}. The elastic
and inelastic processes were evaluated separately and then combined weighted by
the appropriate cross sections. Table~\ref{tab:d200} contains a summary of the
evaluation of the theoretical asymmetry, and the experimental result, broken down
by detector type.

\begin{figure}
\begin{center}
\includegraphics[width=10cm]{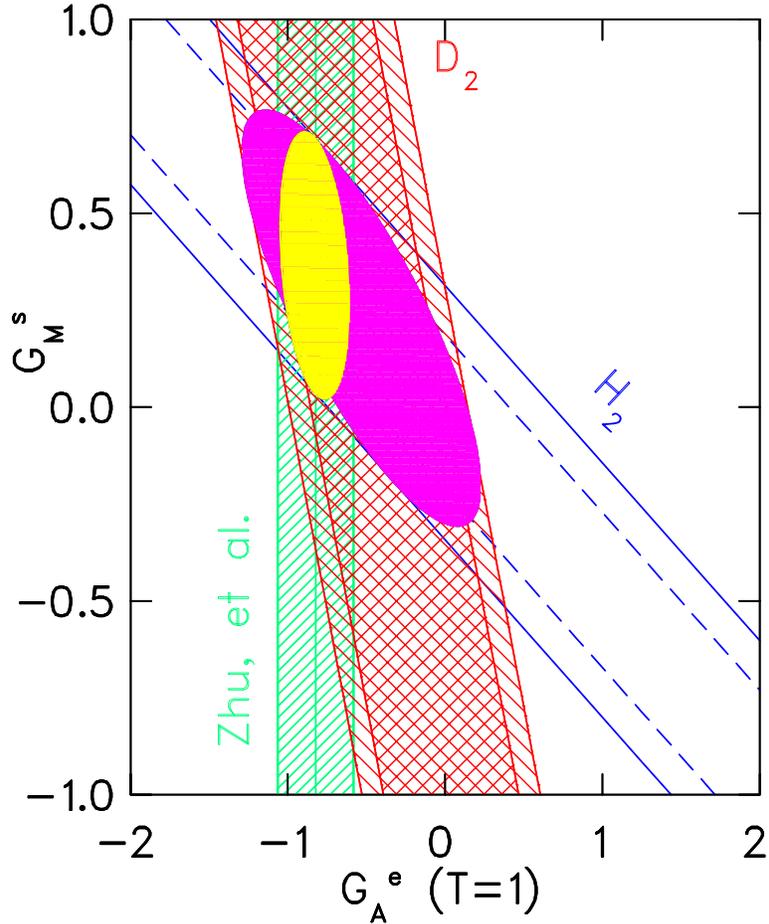}
\caption{Results from the 200~MeV {\sc SAMPLE} data, in the space of
$G_M^s$ vs.~${G_A^e}^{(T=1)}$, along with the theoretically expected
value of ${G_A^e}^{(T=1)}$, using~\protect{\cite{Zhu00}} for the weak
radiative corrections. The ellipses correspond to a 1-$\sigma$ overlap
of the two data sets (larger) and the hydrogen data and theory (smaller).}
\label{fig:gms_gae}
\end{center}
\end{figure}

The resulting theoretical asymmetry is
\begin{equation}
A_d = -7.06 +  0.72 G_M^s + 1.66{G_A^e}^{(T=1)} \, .
\end{equation}
In Figure~\ref{fig:gms_gae} are shown three bands in the space of $G_M^s$
vs.~${G_A^e}^{(T=1)}$ for the hydrogen and deuterium data as well as the
prediction of Zhu~{\it et al}, along with two ellipses representing to 1-$\sigma$
the overlap of the two measurements and also the hydrogen data with the
theoretical band. The resulting values of the form factors taken solely
from the two data sets are
\begin{eqnarray}
G_M^s &=& 0.23 \pm 0.36 \pm 0.40 \nonumber \\
{G_A^e}^{(T=1)} &=& -0.53 \pm 0.57 \pm 0.50 \, ,
\end{eqnarray}
in good agreement with~\cite{Zhu00} and with the determination of $G_M^s$ from
the hydrogen data alone.

\begin{table}
\begin{center}
\begin{tabular}{|c|c|c|c|c|c|c|}
\hline
Det.~\# & $\langle\theta \rangle$ & $A_0$ & $A_1$ & $A_2$ & rel. weight & $A_{meas}$ \\
\hline
1,3,8,10 & 143 & -7.04 &  0.76 & 1.65 & 0.49 & -8.37$\pm$1.03 \\
2,9 & 152 & -7.66  &  0.83 & 1.77 & 0.16 & -10.13$\pm$1.85\\
4,7 & 136 & -6.37  &  0.73 & 1.55 & 0.26 & -6.21$\pm$1.43\\
5,6 & 158 & -8.09  &  0.88 & 1.85 & 0.07 &  -4.99$\pm$2.36\\
\hline
\end{tabular}
\caption{Summary of experimental results and theoretical expectation for the asymmetry,
in ppm, broken down by detector type for the SAMPLE 200~MeV deuterium data.
The quantities $A_0$, $A_1$ and $A_2$ are the leading term and the coefficients
multiplying $G_M^s$ and ${G_A^e}^{(T=1)}$, respectively. The relative weights
were determined by the uncertainty in the measured asymmetries for each detector
type.}
\label{tab:d200}
\end{center}
\end{table}

\begin{figure}
\begin{center}
\includegraphics[width=10cm]{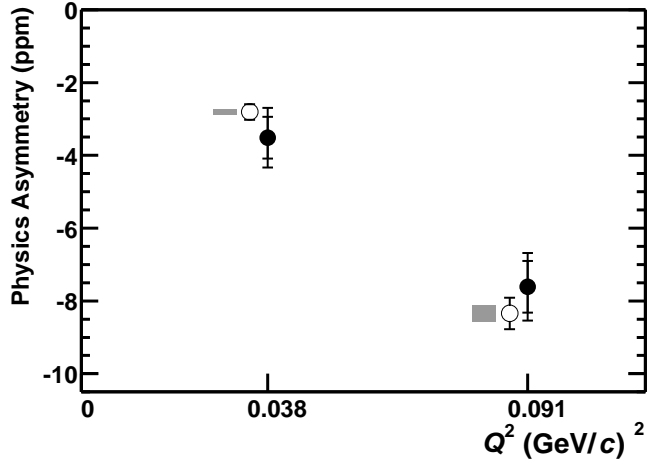}
\caption{Comparison of the two {\sc SAMPLE} deuterium measurements (solid circles) \
to expectation from theory (open circles) using axial radiative corrections as
computed by~\protect{\cite{Zhu00}}. The theory also assumes a value of $G_M^s$
of 0.15~nuclear magnetons. The grey bands represents a change in $G_M^s$ of 0.6~n.m.}
\label{fig:ad_theory}
\end{center}
\end{figure}

In order to provide another experimental determination of ${G_A^e}^{(T=1)}$,
a third {\sc SAMPLE} measurement was carried out at a lower beam energy of 125~MeV,
corresponding to a momentum transfer of 0.04~(GeV/c)$^2$. The experimental
method and apparatus were identical to the 200~MeV measurements, as was the
procedure for determining the dilution corrections and the physics asymmetry.
Three additional systematic uncertainties were required in order to account
for effects seen in this measurement not present in the earlier ones. First,
a nonzero helicity correlation in the beam transmission, caused by
differential beam scraping and an energy-defining slit in the accelerator,
was seen in one of the two half-wave plate states. As a result beam transmission
was used as a parameter in the linear regression to remove helicity correlated beam
effects. Because this correction was larger in one half-wave plate state than the
other (0.14~ppm vs.~0.03~ppm prior to application of dilution corrections),
the relative uncertainty in the correction was also correspondingly larger
(11.2\% vs.~2\%). The uncertainty in the corrections procedure was determined by
comparing the size of the correction using the dependence of the detector yield
on unpolarized beam properties with that using the dependence of the detector
asymmetry on beam helicity correlations. Secondly,
the forward angle luminosity monitors had a nonzero asymmetry even after the
corrections procedure was applied. While in principle the main detector and luminosity
monitor signals should be uncorrelated, a nonzero luminosity asymmetry could
be indicative of an unaccounted for helicity correlated beam property. The
magnitude of potential false asymmetry
in the {\v C}erenkov detector was estimated by multiplying the luminosity
monitor asymmetry with the correlation between the {\v C}erenkov detector
asymmetry and the luminosity monitor asymmetry, resulting in an additional
20\% systematic error. Finally, while again the average background asymmetry was
consistent with zero, the $\phi$-dependence in the background
asymmetry that was present in the 200~MeV hydrogen data (but absent from
the 200~MeV deuterium data) was again seen. The asymmetry was measured with
dedicated runs by placing a set of scintillators in front of the phototube shutters
to enhance the background component, and the background asymmetry was subtracted
from the {\sc OPEN} asymmetry. The systematic uncertainty associated with this procedure
was estimated to be 5\%. The resulting measured asymmetry was
\begin{equation}
A(Q^2=0.038) = -3.51 \pm 0.57 \pm 0.58\, {\mathrm {ppm}} \, .
\end{equation}
The theoretical value to which the measurement should be compared was computed as
described above, resulting in
\begin{equation}
A_d = -2.14 + 0.27 G_M^s + 0.76 {G_A^e}^{(T=1)}\, {\mathrm {ppm}} \, .
\end{equation}
While no corresponding hydrogen data were taken at the lower beam energy, the
two deuterium measurements can be compared to theoretically expected values
assuming a particular value of $G_M^s$ at the lower momentum transfer of
$Q^2=0.038$~(GeV/c)$^2$. The dependence of the deuterium asymmetry on $G_M^s$ is,
however, very weak. A comparison between the data and theoretical
expectation assuming a constant value
of $G_M^s=0.15$ is shown in Figure~\ref{fig:ad_theory}. The boxes represent
the variation in the theory for a
change in $G_M^s$ of 0.6. The good agreement indicates a weak
$Q^2$ dependence of ${G_A^e}^{(T=1)}$, consistent with expectation.

\begin{figure}
\begin{center}
\includegraphics[width=10cm]{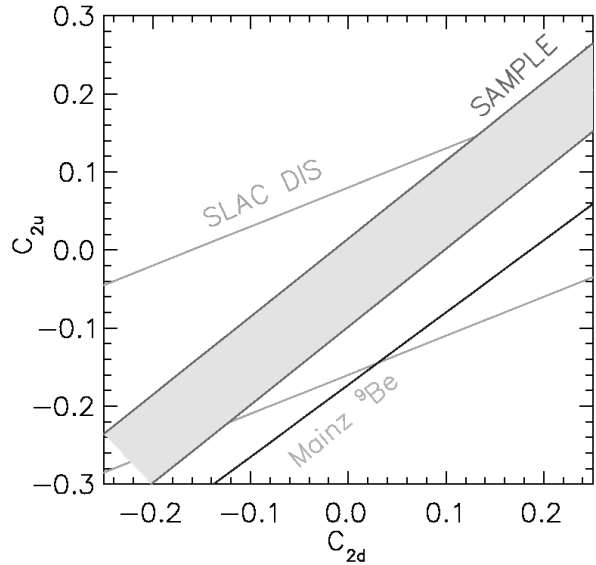}
\caption{Existing measurements of $C_{2u}$ and $C_{2d}$, shown as 1-$\sigma$ uncertainty bands. Only the
200~MeV {\sc SAMPLE} result is shown. The prediction for these quantities in the Standard Model, using the
global best fit values of the masses of the top quark and Higgs boson and the strong and electromagnetic
coupling constants (see ~\protect\cite{PDG03}), is $C_{2u}=-0.0360$, $C_{2d}=0.0265$.} \label{fig:c2u_c2d}
\end{center}
\end{figure}

Two additional remarks are in order regarding the {\sc SAMPLE}
experimental measurements.  Assuming that a determination of nucleon
form factors can ultimately be related to electron quark couplings,
the axial form factor measurements in the two deuterium experiments
can be recast in terms of the parameters $C_{2u}$ and $C_{2d}$ that
represent electron-quark axial couplings. Prior to the {\sc SAMPLE}
measurements the only experimental limits on $C_{2u}$ and $C_{2d}$
were from the first parity-violation electron scattering experiment to be
performed~\cite{Pre79} and from the parity-violating quasielastic
scattering measurement carried out at the Mainz microtron in the
mid-1980's~\cite{Hei89}. The two {\sc SAMPLE} deuterium measurements
are sensitive to the combination
\begin{equation}
C_{2u}-C_{2d} = -\frac{{G_A^e}^{(T=1)}\left(1-4\s2w\right)}{G_A(Q^2)} \, .
\end{equation}
It is modified by the 1-quark radiative corrections, and,
in the case of elastic electron-nucleon scattering, it is also modified by the somewhat
more uncertain multi-quark corrections discussed in~\cite{Zhu00}. In order
to compare directly to the deep-inelastic scattering data, the multi-quark
corrections must be removed from the SAMPLE data. While this is small
contribution, it dominates the uncertainty in the radiative corrections.
The one-quark radiatively
corrected values of the couplings within the context of the Standard Model
are $C_{2u}=-0.0360$ and $C_{2d}=0.0265$. Extrapolating the neutral
weak axial form factor $G_A(Q^2)$ using the dipole parameterization described in
the introduction, and removing the multi-quark corrections results in the following
from the {\sc SAMPLE} deuterium 200~MeV and 125~MeV
data sets, respectively.
\begin{eqnarray}
C_{2u}-C_{2d} &=& -0.042\pm 0.040 \pm 0.035 \pm 0.02  \nonumber \\
C_{2u}-C_{2d} &=& -0.12\pm 0.05 \pm 0.05 \pm 0.02 \pm 0.01 \\
\end{eqnarray}
where third uncertainty is that due to the multi-quark radiative corrections.
For the 125~MeV result, the last uncertainty corresponds to
variation of $G_M^s$ by $\pm 0.6$, because it is undetermined at this
momentum transfer. These values are in good agreement with the Standard Model prediction,
and the improvement over the earlier measurements is shown in Figure~\ref{fig:c2u_c2d}.

Another byproduct of the {\sc SAMPLE} experiment was the first measurement of the beam spin
asymmetry on a nucleon using a transversely polarized beam~\cite{Wel01}
(also referred to as a vector analyzing power). Scattering from a transversely polarized
beam and an unpolarized target produces a parity conserving asymmetry for electrons scattered in
opposite directions about the polarization axis. With an azimuthally symmetric detector
the asymmetry can be seen as a $\cos\phi$ dependence in the counting rate asymmetry when
the beam polarization direction is reversed. It is a time-reversal odd observable
that, to lowest order, can only arise from two-photon exchange. Recently there has been renewed
interest in two photon processes for two reasons. As discussed above, it
has recently been determined that two-photon exchange processes can be an
important contribution to the determination of the proton's electric form factor from
unpolarized cross section data. They are also interesting as a probe of nucleon
generalized polarizabilities, through virtual Compton Scattering (in which a real
photon is emitted).

\begin{figure}
\begin{center}
\includegraphics[width=10cm]{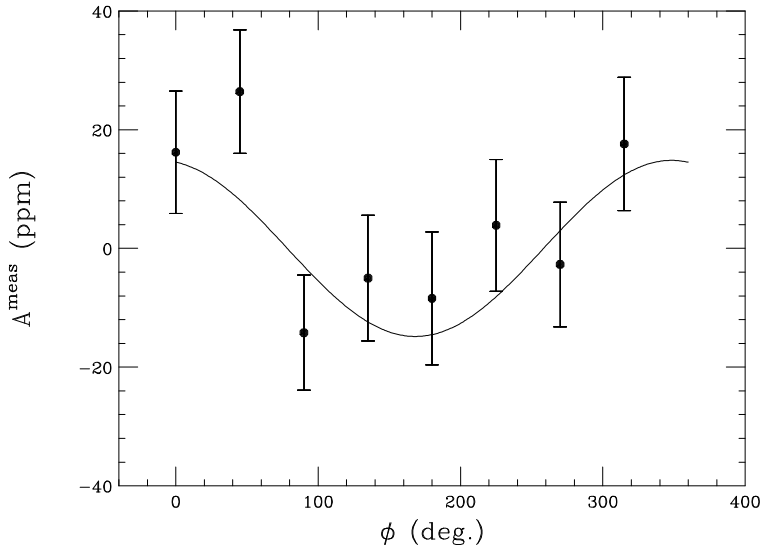}
\caption{Asymmetry as a function of azimuthal angle $\phi$ for transversely
polarized beam, as measured in the {\sc SAMPLE} experiment. The curve is the best fit
to a set of four combined measurements.}
\label{fig:vector-sample.ps}
\end{center}
\end{figure}

In each of the {\sc SAMPLE} measurements, a few days of data taking
were spent with the beam polarization aligned transversely, both in
and out of plane. The measured asymmetry as a function of azimuthal
angle is shown in Figure~\ref{fig:vector-sample.ps}, along with a fit
to the data. The $\phi$ dependence in the data can clearly be seen,
the amplitude representing the resulting beam spin asymmetry, found to
be $A_T = -16.4\pm 5.9$~ppm, about a factor of two larger than
predicted by the calculation of Afanasev~{\it et al.}~\cite{Afa02},
when only ground state protons are considered in the intermediate
state between the two photons. Calculations of the beam spin asymmetry
were also carried out at higher energies~\cite{VDH03}, where proton
excitations in the intermediate state are expected to be significant,
at the kinematics appropriate for the Mainz PVA4 experiment.  Two results
from the PVA4 experiment have recently been reported~\cite{Bau04}: at
$E=0.855$~GeV (and $\theta_e=35^\circ$), $A_T=-7.62\pm 2.34 \pm 0.80$~ppm
and at $E=0.569$~GeV, $-8.28\pm 0.93\pm 0.49$~ppm.
The authors of~\cite{VDH03} speculate that contributions from threshold
meson production, which were neglected in their first calculation, could
cause an additional enhancement, and a new calculation that uses the MAID
unitary isobar model for the intermediate states has recently become
available~\cite{Pas04}. Additional measurements of $A_T$ at
higher momentum transfer are anticipated in the future from the
G$^0$ and {\sc HAPPEX} experiments, where simple nucleon excitation in
the intermediate state may no longer be an adequate description, but at
energies below that appropriate for a parton model description.

\section{Other Experiments}
\label{sec:compare}

\subsection{HAPPEX at Jefferson Laboratory}
\label{sec:happex}

The first measurements of parity-violating electron scattering at Jefferson
Laboratory~\cite{Ani00}
were carried out by the HAPPEX collaboration, who used a 3.3~GeV
polarized beam on a 15~cm hydrogen target and detected the scattered electrons
using the pair of high resolution spectrometers (HRS) in Hall A at 12.5$^\circ$.
The measured asymmetry, at $Q^2$=0.48~(GeV/c)$^2$, is sensitive to the combination
$G_E^s + 0.39G_M^s$. With a counting rate of approximately 1~MHz per spectrometer,
the spectrum of scattered electrons was integrated over a 30~ms window. A set
of Pb-scintillator total absorption counters was used instead of the standard HRS
tracking detector package, but the hardware resolution of the spectrometers was
sufficient to spatially separate elastic from inelastic scattering events.
Custom electronics with 16-bit analog-to-digital converters and low differential
nonlinearity were developed for additional optimization of the detection scheme.
The helicity of the beam was flipped at a rate of 30~Hz, with the first of
a pair being chosen randomly and the second being chosen as the complement of
the first, allowing measurement of the PV asymmetry at a rate of 15~Hz.
A half wave plate was either inserted or removed approximately once per day,
allowing manual reversal of the beam helicity without changing the electronics.
The beam polarization was measured with a combination of a Mott polarimeter
near the polarized injector region, a M{\o}ller polarimeter on the Hall A
beam line, and, for the later running, a Compton polarimeter that operated
concurrently with normal data taking.  Additional detail and a recent review
of the HAPPEX experimental techniques can be found in~\cite{Kum00}.

The HAPPEX experiment was the first to make use of a strained GaAs photocathode in the
polarized electron source for parity violation experiments. The
strained cathode allows for the possibility of significantly higher beam
polarization, in excess of 70\%, but with a quantum efficiency that is lower
than normal bulk GaAs by about an order of magnitude. Improvements in laser
power over the years, along with the better match between the time distribution
of the laser light and the needs for the JLab beam made it possible for
HAPPEX to take advantage of the increased polarization. One complication, as discussed
above, is that such cathodes usually have an intrinsic analyzing power for
linear polarization in the incident laser light that
could potentially result in significant helicity correlated position differences
of the beam on the experimental target. These effects were kept to
a negligible level by insertion of a rotatable half-wave plate in the
laser beam and with a feedback system nulling any helicity-correlated intensity
asymmetry. This feedback system has continued to improve for the next generation
of experiments. New developments in crystal fabrication may eliminate this
complication in the future.

\begin{figure}
\begin{center}
\includegraphics[angle=0,width=10cm]{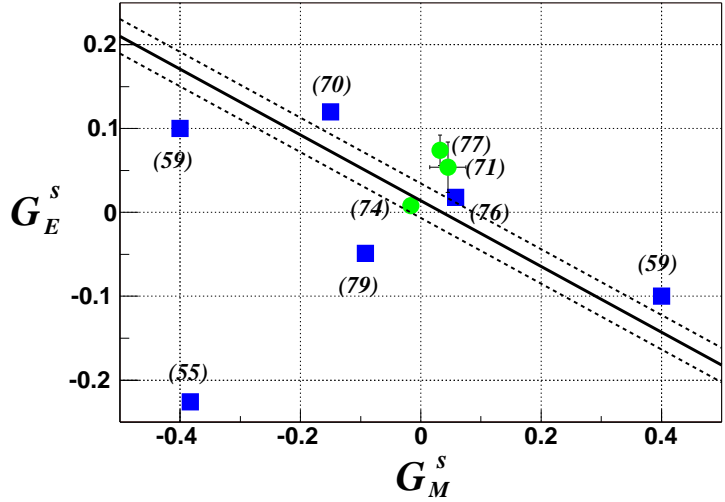}
\caption{The band shows the allowed region of $G_M^s$ and $G_E^s$ at $Q^2$=0.48~(GeV/c)$^2$ coming from two
completed HAPPEX experiments~\protect{\cite{Ani00}}, along with several models. The figure
is taken with permission from~\protect{\cite{Ani04}}, with the reference numbers modified to reflect
those in this work.  The squares are calculations done prior to the measurement, circles subsequent.}
\label{fig:HappexI}
\end{center}
\end{figure}

The experimentally determined asymmetry from the two HAPPEX runs combined is
$A_{exp}=-15.05\pm 0.98\pm 0.56$~ppm, corresponding to $G_E^s+0.39G_M^s =
0.025 \pm 0.020 \pm 0.014$ where the last uncertainty is due to knowledge
of the nucleon EM form factors. HAPPEX thus precludes most of
the parameter space in which $G_E^s$ and $G_M^s$ have the same sign at
$Q^2=0.48$~(GeV/c)$^2$, as shown in Figure~\ref{fig:HappexI}.

The future program for HAPPEX includes a forward angle measurement at
$Q^2$=0.1 (GeV/c)$^2$ on hydrogen~\cite{Kum99}, as well as the first measurement of
the PV asymmetry in elastic electron scattering from helium~\cite{Arm00}.
Due to the fact that $^4$He is a spin-0, isospin-0, target, only a single weak
form factor exists and it can be directly related to $G_E^s$ with a good
model of the $^4$He nucleus. Theoretical expectations are that contributions
to the asymmetry from many-body effects in the helium are
negligible~\cite{MusD93} at low momentum transfer. The combined measurements,
or the new hydrogen measurement combined with the SAMPLE result, will
result in a determination of the electric strangeness radius $r_s^2$.

\subsection{PVA4 at Mainz}
\label{sec:pva4}

\begin{figure}
\begin{center}
\includegraphics[width=10cm]{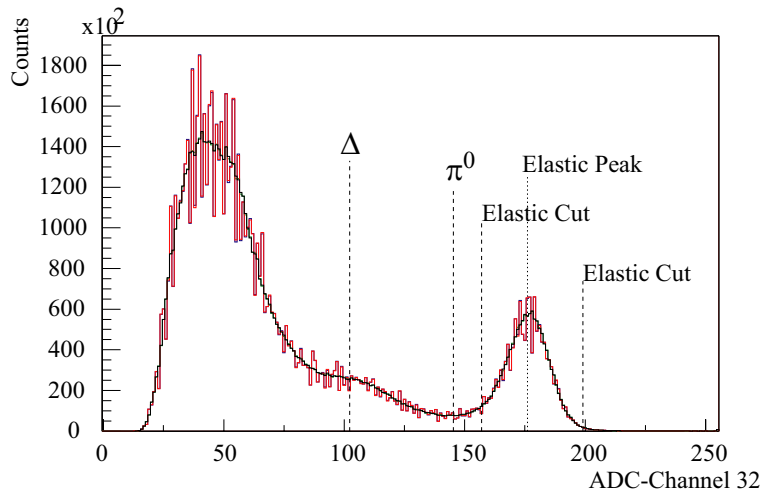}
\caption{Real-time histogram of the energy spectrum of detected particles in the PbF$_2$ detector used in the
PVA4 experiment at Mainz. The red histogram is the raw data,
overlaid with (black line) the spectrum after corrections for differential
non-linearity in the analog-to-digital converter.} \label{fig:PVA4-energy}
\end{center}
\end{figure}

The PVA4 collaboration at Mainz has taken a different experimental
approach~\cite{Maa00}.  The custom detector system has no magnetic field
and azimuthal symmetry as in the case of SAMPLE, but is optimized for forward
angle measurement. The detector has sufficient segmentation and specialized electronics so
that counting the scattered particles is feasible despite very high
scattered electron rates. The scattered electrons
are detected over a range of angles centered at 35$^\circ$ with
a PbF$_2$ {\v C}erenkov shower calorimeter.
The calorimeter design consists of 1022 PbF$_2$ crystals of 16-20 radiation lengths
thickness, arranged in 7 rings, and processed in 3x3 modules with self-triggering and
histogramming electronics to collect in real time an energy spectrum of the scattered particles.
An energy deposition above a specified threshold triggers the digitization of
the summed output of the 3x3 cluster of crystals, in which the charge is integrated
over 20~ns and histogrammed. The energy resolution of the detectors must be sufficient
to separate the 10~MHz of elastically scattered electrons from the 90~MHz of
inelastic electrons coming from threshold pion and resonance production. The achieved
energy resolution was 3.5\%/$\sqrt{E}$~\cite{Wie03} for 1~GeV particles.
A typical energy spectrum of the detected events is shown in Figure~\ref{fig:PVA4-energy}.

The first PVA4 measurement~\cite{Maa04} was at a beam energy of 855~MeV, corresponding to
$Q^2$=0.23~(GeV/c)$^2$ and a sensitivity to the combination $G_E^s+0.22 G_M^s$, in
which half of the detector was instrumented. A 20~$\mu$A beam of polarized electrons
was incident on a 10~cm liquid hydrogen target. The
beam-target luminosity was monitored with a set of eight water {\v C}erenkov detectors
placed symmetrically about the beam line. The beam polarization was approximately 80\%, from
a strained GaAs in the polarized electron source. The beam helicity was reversed every 20~ms,
with an additional manual slow reversal by periodic insertion of an additional
half-wave plate. Potential helicity-correlated asymmetries in the incident beam current were
kept below several parts per million by adjusting the angle of another, permanently
installed, half-wave plate.  Beam position and angle incident on the target were
measured with microwave cavity monitors, and beam polarization was measured periodically
with a M{\o}ller polarimeter on a different beam line.

The experimental asymmetry after all dilution corrections was measured to be
$A_{exp} = -5.44 \pm 0.54 \pm 0.26$~ppm~\cite{Maa04}, with an expectation,
assuming no strange quark effects, of $A_0 = -6.3\pm 0.43$~ppm, where the uncertainty
in the theoretical value arises predominantly from uncertainties in the neutron
electromagnetic form factors. This difference of approximately 1$\sigma$
hints that $G_E^s$ and $G_M^s$ are either both small or have
opposite sign. Additional data have already been taken, the run concluding in June 2003,
at 570~MeV beam energy, corresponding to $Q^2$=0.1~(GeV/c)$^2$.
Combining these data with the results from SAMPLE will allow the first
explicit experimental limits on $G_E^s$. Future plans involve reversing the orientation
of the detector such that scattered electrons can be detected at 140$^\circ$-150$^\circ$
at $Q^2$=0.23 and 0.48~(GeV/c)$^2$, which, when combined with
the existing HAPPEX and PVA4 data would allow separate determination of $G_E^s$ and
$G_M^s$ over a range of momentum transfer.

\begin{figure}
\begin{center}
\includegraphics[width=10cm]{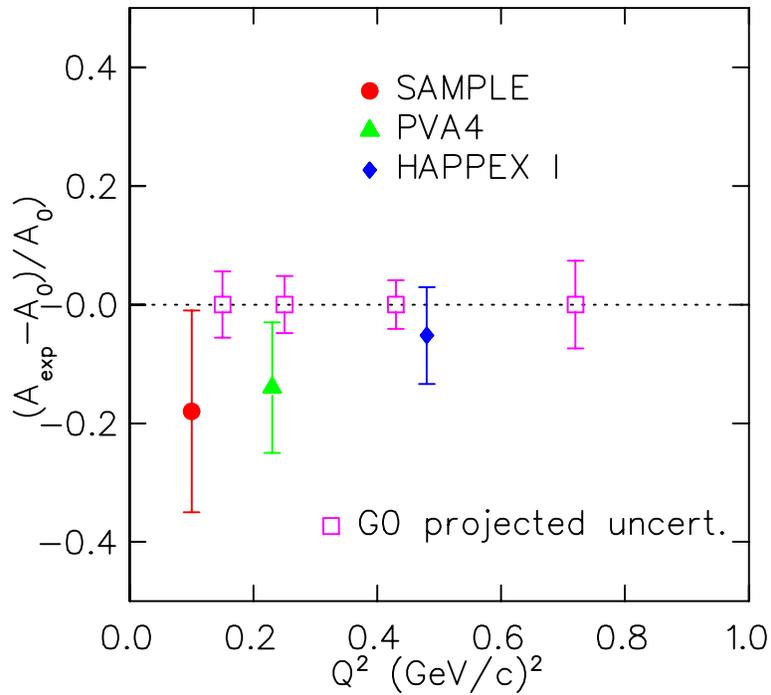}
\caption{Summary of the available data on parity-violating $e$-$p$ scattering, plotted as
the fractional deviation of the measured asymmetry from the expectation with no strange quark
contributions. The HAPPEX point is from~\cite{Ani04}, the PVA4 point from~\cite{Maa04}, the
SAMPLE point from~\cite{Spa03}, and the G$^0$ points are projections of the precision expected from
the forward angle running based on information from the 2004 engineering run.}
\label{fig:asym-dev}
\end{center}
\end{figure}

A summary of the three available measurements of the parity-violating asymmetry in elastic $e$-$p$
scattering is shown in Figure~\ref{fig:asym-dev}, as the deviation of the asymmetry from that
expected with no strange quark contributions. While each of the individual measurements is consistent with
zero at the 1$\sigma$-level, the three measurements taken together hint at a possible nonzero
contribution. The data set is at present insufficient to distinguish between
no strange quark contributions and cancellations between $G_E^s$ and $G_M^s$.  The next generation
of experiments, which include the future programs of HAPPEX and PVA4 as well as the G$^0$ experiment
described below, will address this problem.

\subsection{The G$^0$ Experiment at JLab}
\label{sec:G0}

\begin{figure}
\begin{center}
\includegraphics[width=10cm]{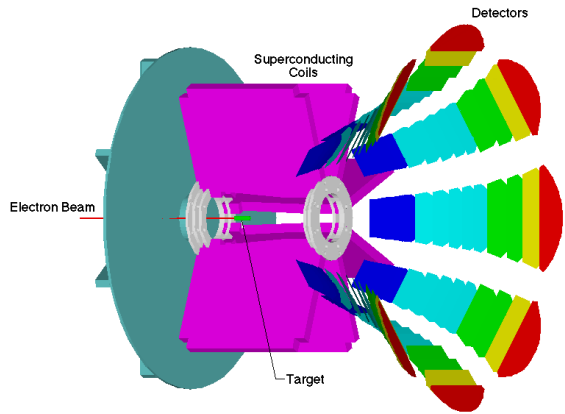}
\caption{Schematic of the G$^0$ apparatus used in its forward-angle configuration. Each
detector color corresponds to a different value of momentum transfer $Q^2$, ranging from
0.16~(GeV/c)$^2$ to 0.9~(GeV/c)$^2$.}
\label{fig:G0-schematic}
\end{center}
\end{figure}

\begin{figure}
\begin{center}
\includegraphics[width=10cm]{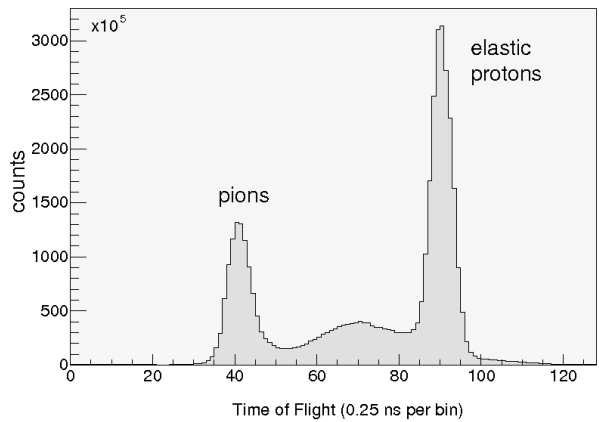}
\caption{Time of flight histogram, as accumulated in a 33~msec helicity window in the
French flash-TDC-based electronics. The left peak is due to photo-produced pions, in
the right peak are elastically scattered protons, and the broad distribution in between
results from inelastically scattered protons.}
\label{fig:g0-tof}
\end{center}
\end{figure}

The G$^0$ experiment at JLab is a dedicated apparatus designed to
determine $G_E^s$, $G_M^s$ and $G_A^e$ from a single experimental
apparatus over a broad $Q^2$ range.  The detector consists of a
superconducting toroidal spectrometer with an array of scintillators
along the focal plane to determine the PV asymmetry at both forward
and backward scattered electron angles.  Polarized electrons are
scattered from a 20~cm liquid hydrogen target.  In the forward
configuration, the recoil protons are detected and sorted by $Q^2$
covering the range $0.1 < Q^2 < 1.0$~(GeV/c)$^2$. A schematic of the
apparatus is shown in Figure~\ref{fig:G0-schematic}. The flight time
from the target to the scintillator array is determined for each
scattered particle, allowing the rejection of photoproduced pions and
many of the protons generated from inelastic processes.  In order to
determine the flight times of the detected particles, which range from
5-25~ns, the time structure of the JLab beam is modified from its
nominal 2~ns between micropulses to be 32~ns between micropulses.  The
helicity of the beam is selected pseudo-randomly in a pattern of four
``windows'', or quartets, either ``$+--+$'' or ``$-++-$'', such that a
new window is selected every 33~ms.~\symbolfootnote[2]{The choice of a quartet sequence
rather than the ``pair'' sequence used in other experiments
reduces the sensitivity to short-term linear drifts in beam properties.}
The detected rate approaches
1~MHz, so the flight times are stored electronically and read out as
histogrammed data on each helicity reversal. Two complementary sets of
electronics are used to read out the detectors: for the odd (North
American) octants the arrival time of the events is digitized and then
sent to a set of scalers, from which the time-of-flight histogram is
reconstructed with a resolution of 1~nsec per time bin. For the
even-numbered (French) octants, custom modules were constructed with
flash TDCs to digitize and stored the time spectrum, with a resolution
of 0.25~nsec per bin, for each 33 ms window.  In
Figure~\ref{fig:g0-tof} is shown a typical time spectrum of the
scattered events. The inelastically scattered protons that cross under
the elastic proton peak are the primary source of background; they come
primarily from scattering in the aluminum windows of the target and
from pion photoproduction in the liquid hydrogen.  The measured
asymmetry must be corrected both for any dilution factor and for any
potential parity-violating asymmetry associated with these
processes. Because this is a counting experiment, the detected rate
must also be corrected for electronic deadtime, which has been
measured to be 10-15\%, and which can produce a false asymmetry if
there is a helicity correlated change in the incident beam
current. Through a series of tests this correction has been
demonstrated to be under control at a level that would produce a
negligible false asymmetry in the data. Finally, the data must be
corrected for any additional helicity correlations in the position and
angle of the incident beam. As discussed in Section~\ref{sec:beam}
above, such corrections can be minimized by feedback systems on the
laser light that produces the polarized electron beam.

The G$^0$ collaboration has recently completed its second engineering run and
will begin production data taking in mid-March 2004. During this run the asymmetry
in the background processes was measured through a combination of empty target
and solid target running.

While a definitive determination of the relative signs and/or magnitude of $G_M^s$ and $G_E^s$ will
require the backward angle measurements as well, this first set of data will both extend
the kinematic reach and improve the precision of the data shown in Figure~\ref{fig:asym-dev}.

In order to separately determine $G_E^s$, $G_M^s$ and $G_A^e$, three
independent measurements of parity-violating electron scattering are
required. The G$^0$ experiment will accomplish this by an independent set
of measurements with the apparatus reversed such that scattered
electrons can be detected in the backward direction at
108$^\circ$. Measurements in this configuration are planned with both
hydrogen and deuterium targets to provide the two additional constraints.
Three sets of measurements, each requiring a different
magnetic field setting, are planned, corresponding to $Q^2$ = 0.25,
0.5 and 0.8~(GeV/c)$^2$. The scattered electron rate is significantly
lower, but is now spread over the entire detector array. Inelastically
scattered electrons are discriminated from the elastic electrons of
interest with an auxiliary set of detectors at the exit of the magnet
cryostat that constrain the tracks of the particles leaving the
magnet. The negatively charged pion rate from liquid hydrogen, which
can only arise from two-step processes, is low, and thus does not
present a significant background, but from liquid deuterium the rate
is expected to be appreciable, so a set of aerogel {\v C}erenkov veto
counters close to the Cryostat exit detectors will also be added.  New
custom electronics will identify and sort the elastically scattered
events through a coincidence matrix selection. It should be noted that
the inelastically scattered electrons, which are primarily due to
excitation of the $\Delta$ resonance, will be available for analysis
as well~\cite{Wel00}, thus offering an opportunity to determine the
$N$-$\Delta$ axial transition form factor via the weak neutral current
process.

\section{Future Directions}
\label{sec:future}

\subsection{Standard Model Tests using Parity-Violating Electron-Nucleon
Scattering}
\label{sec:ew-tests}

The primary focus of this review article is the use of parity-violating electron scattering on the nucleon as
a probe of nucleon structure.  In the formalism used to extract the strange vector form factors from these
experiments, it is assumed that the electroweak Standard Model couplings of the neutral weak $Z$-boson to the
quarks summarized in Table~\ref{tab:charges} are correct.  Given the precision with which the strange form
factors are anticipated to be determined, this is a reasonable assumption. Historically, the first
experimental measurements of parity-violating electron-nucleon scattering of Prescott~{\it et
al.}~\cite{Pre79} were measurements of the electroweak quark couplings at a time when the electroweak Standard
Model was in its infancy.  Since that time, the electroweak Standard Model has been verified to high
precision.  Its success makes the current set of strange vector form factor measurements possible. With the
increased precision now possible in parity-violating electron-nucleon scattering, experiments are being
developed to again test the validity of the electroweak Standard Model, thus returning the field to its
historical origins.  Several new experiments are currently being
developed for this purpose.

To motivate the need for new Standard Model tests using parity-violating
electron-nucleon scattering, we first
briefly review the relevant history in this area.  The 1978 experiment of
Prescott~{\it et al.}~\cite{Pre79}
at SLAC was a measurement of the parity-violating asymmetry in the deep
inelastic scattering (DIS) of
longitudinally polarized electrons on deuterium.  The experiment used incident electron
beam energies in the range of 16-22 GeV.  This experiment yielded a determination of the weak mixing
angle, $\sin^2 \theta_W = 0.224\pm 0.020$, which
was one of the most accurate determinations at that time.  The existence of the neutral weak
current had been established earlier in neutrino experiments in 1973, but the
extraction of a consistent value of the weak mixing angle from another process was an
important confirmation of the validity of the electroweak Standard Model.

In the 1980's, a large amount of experimental evidence in support of the
electroweak Standard Model was accumulated.  The most precise
measurements of electroweak processes were made in direct production
of the $Z$-boson at SLAC's SLC and CERN's LEP $e^+e^-$ colliders.
These led to a very accurate and internally self-consistent determination of the
weak mixing angle, $\sin^2 \theta_W$, at the $Z$-pole ($Q^2 = M^2_Z$).
However, relatively few measurements of $\sin^2 \theta_W$ have been made
away from the $Z$-pole. Experiments that probe the value of $\sin^2 \theta_W$
far from this value of momentum transfer may reveal physical processes
which do not couple strongly to the $Z$ and are outside of the current
realm of the Standard Model.
As a result, such experiments have an important
role in the search for extensions to the Standard Model. Two programs that have
been pursued are measurement of parity-violation in atomic systems~\cite{Ben99}
and of parity-violation in electron-electron (M{\o}ller) scattering~\cite{Ant03}.
In new parity-violating electron-nucleon scattering experiments,
precision measurements of $\sin^2 \theta_W$
well below the $Z$-pole ($Q^2 << M^2_Z$) are proposed.  To make a meaningful
statement, the new generation of parity-violating Standard
Model tests anticipate relative uncertainties on $\sin^2 \theta_W$
in the $\sim 0.3 - 0.7\%$ range, in comparison to the $9\%$
measurement of the original SLAC $e$-$d$ DIS experiment.
The advances in polarized electron beam delivery, beam feedback
and control, and high power cryotarget technologies appear to make this
possible.

\subsubsection{The $Q_{weak}$ Experiment at Jefferson Laboratory}
\label{sec:qweak}

\begin{figure}
\begin{center}
\includegraphics[width=8cm,angle=90]{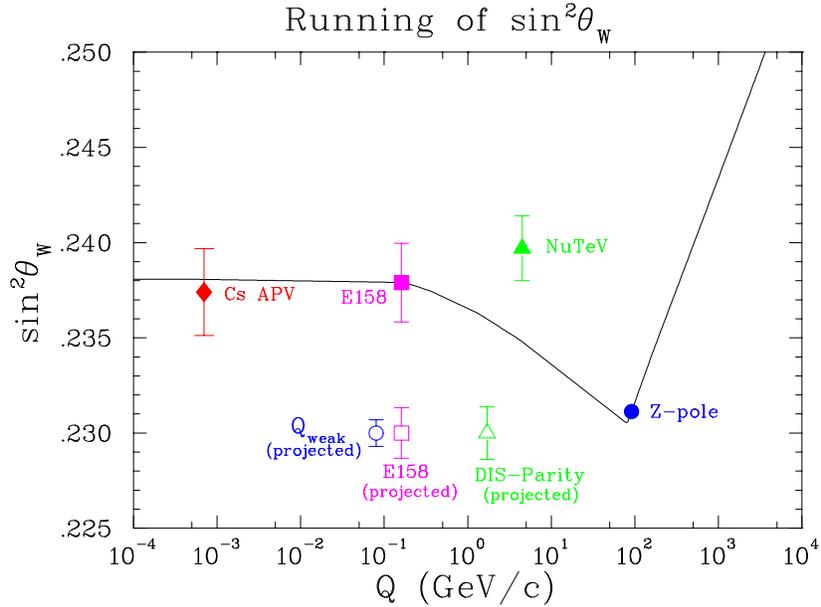}
\caption{ The calculated running of the weak mixing angle in the Standard Model, defined in the modified
minimal subtraction scheme~\protect{\cite{Erl03}}, is shown as the solid line. The solid data points are from
published results: atomic parity violation(APV)~\protect{\cite{Ben99}}, parity-violating M{\o}ller scattering
(E158)~\protect{\cite{Ant03}}, deep inelastic neutrino-nucleus scattering (NuTeV)~\protect{\cite{Zel02}}, and
the world average of $Z$-pole measurements at LEP and SLC.  The open symbols are arbitrarily placed on the
vertical axis, and they represent proposed experiments: $Q_{weak}$~\protect{\cite{Car01}}, the anticipated
reduction in the E158 error bar after further data taking, and the DIS parity experiments described in
section~\ref{sec:disparity}.}
\label{fig:runs2w}
\end{center}
\end{figure}

The $Q_{weak}$ experiment at Jefferson Laboratory~\cite{Car01} will be
a measurement of the proton's neutral
weak charge using parity-violating electron-nucleon scattering.
To lowest order in the Standard Model, the proton's neutral weak
charge is the sum of the neutral weak charges of its valence quarks:
\begin{equation}
Q^p_{W} = G^{Z,p}_E(Q^2=0) = 2 q^{u}_{W} +  q^{d}_{W} = 1-4 \sin^2 \theta_W.
\end{equation}
The proposed measurement will yield a precise value of
$\sin^2 \theta_W$ at $Q^2 \sim 0.03 \ \hbox{GeV}^2$.
In the Standard Model, the effective value of
$\sin^2 \theta_W$ is predicted to vary with $Q^2$, similarly to
the running of the coupling constants of QCD and QED.  To date,
however, the predicted ``running of $\sin^2 \theta_W$'' in the electroweak
Standard Model, shown in Figure~\ref{fig:runs2w}, has
not yet been confirmed experimentally.  The predicted
shift is $\Delta \sin^2 \theta_W = +0.007$ at low $Q^2$
with respect to the $Z$-pole best fit value of $0.23113 \pm 0.00015$.
Also shown in Figure~\ref{fig:runs2w} are data
from existing and proposed experiments.  Below the $Z$-pole, there are
three published data points.  The two existing measurements at
low $Q^2$ are from atomic parity violation experiments in the cesium atom~\cite{Ben99},
and from a parity-violating M{\o}ller scattering
experiment at SLAC~\cite{Ant03} (E158).  The E158 collaboration expects
a reduction in their error bar of about a factor of 2 with further data taking.
Finally, the NuTeV point is from $\nu$-nucleus deep inelastic
scattering~\cite{Zel02}.  Of the three points, only the NuTeV result
appears to show a deviation from the Standard Model.  While this
could be a signal for new physics, there are more conventional
explanations that have been advanced to explain this discrepancy
including nuclear effects in the iron target~\cite{Mil02},
QCD effects~\cite{Dav02}, and the possibility of non-identical
strange sea ($s(x)$ and $\overline{s}(x)$) parton distributions as discussed in
Section~\ref{sec:otherobservables}. The anticipated relative uncertainty in
$\sin^2 \theta_W$ from the $Q_{weak}$ experiment is $\sim 0.3\%$, which would
correspond to a 10$\sigma$
deviation from the $Z$-pole value if the running of the weak mixing
angle to low $Q^2$ is in agreement with the Standard Model prediction.
Physics beyond the Standard Model that could enter into $Q^{p}_W$ has
been considered in~\cite{Erl03}, which includes extra neutral weak
bosons, supersymmetry, and leptoquarks.

The $Q_{weak}$ experiment will provide a measurement of the asymmetry in the elastic
scattering of 1.165 GeV longitudinally polarized electrons
from an unpolarized liquid hydrogen target.  In the limit of forward
angle scattering, the asymmetry can be written as:
\begin{equation}
A \sim \left[ \frac{-G_F}{4 \pi \alpha \sqrt{2}}\right]
[ Q^2 Q^p_W+Q^4 B(Q^2)],
\end{equation}
where $Q^p_W$ is the proton's weak charge and $B(Q^2)$ is a hadronic
form factor contribution that involves the proton and neutron
electromagnetic and neutral weak form factors.  The expected data on
neutral weak form factors from the forward angle experiments at higher
$Q^2$ discussed in
section~\ref{sec:compare} is expected to be sufficient to
constrain $B(Q^2)$ so that the uncertainty resulting from this term in the
extraction of $Q^p_W$ will be less than the expected statistical error.
At the proposed $Q^2 = 0.03 \ \hbox{GeV}^2$, the expected asymmetry is
$A \sim -0.3\ \hbox{ppm}$.  A determination of $\sin^2 \theta_W$ at the
desired level of 0.3\% requires a determination of $Q^p_W$ to an
accuracy of $\sim 4\%$ (combined statistical and systematic uncertainties).

\begin{figure}
\begin{center}
\includegraphics[width=14cm]{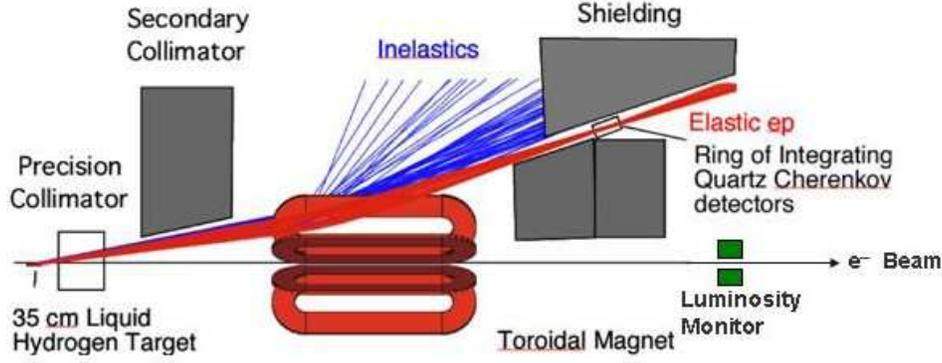}
\caption{Conceptual diagram of the $Q_{weak}$ experimental
apparatus showing the target, collimation system, toroidal magnet,
shielding, detectors, and luminosity monitor.
The elastically scattered electrons of interest are
focused onto the quartz detectors,
while the inelastically scattered electrons are swept away.}
\label{fig:qweak_app}
\end{center}
\end{figure}

A conceptual diagram of the $Q_{weak}$ experimental apparatus is
shown in Figure~\ref{fig:qweak_app}.  The incident $\sim 80\%$
polarized, 180 $\mu$A electron
beam will scatter from a 35 cm liquid hydrogen target.  Scattered electrons
in the range $8\pm2^{\circ}$ are selected by a precision collimator.
A room temperature toroidal magnetic spectrometer focuses elastically
scattered electrons onto a focal plane, while inelastic electrons are
swept into shielding.  The elastically scattered electrons will be
detected in 8 quartz \v Cerenkov bars. Due to the high count rates
($\sim 0.65$ GHz/detector), these detectors will be integrated over
a 33~msec window rather than counting individual events.
A luminosity monitor located at a very forward angle
will be used to monitor for the possible presence of target
density fluctuations and to serve as a sensitive detector for any false
asymmetries.
An auxiliary tracking system
consisting of three sets of tracking chambers will be employed at
low beam currents to accurately measure the average $\langle Q^2 \rangle$ of
the experiment.  High precision electron beam polarimetry will be
carried out with the existing Jefferson Lab Hall C M{\o}ller
polarimeter~\cite{Hau01} and a new Compton polarimeter under construction.
Further details about the $Q_{weak}$ experiment can be found in
\cite{Mit03} and references therein.

\begin{figure}
\begin{center}
\includegraphics[width=8cm]{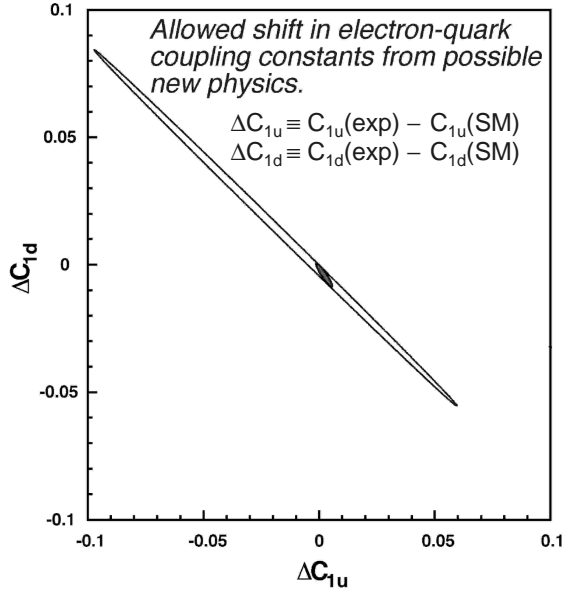}
\caption{Present and anticipated 90\% confidence level
constraints on the possibility of deviations of the electron-quark
vector couplings from the Standard Model values.  The larger
ellipse denotes the present limits derived from atomic parity
violation~\protect{\cite{Ben99}}, parity-violating elastic scattering from
the $^{12}$C nucleus~\protect{\cite{Sou90}}, and the SLAC DIS $e-d$
experiment~\protect{\cite{Pre79}}. The smaller, shaded ellipse shows
the constraints after the inclusion
of a measurement with the expected $Q_{weak}$ precision; the ellipse
is centered on the Standard Model prediction.}
\label{fig:c1_u_d}
\end{center}
\end{figure}

The observables of low energy parity-violating electron-nucleon
scattering experiments
can be analyzed in a model-independent way by using the low
energy effective neutral current electron-quark Lagrangian
given by~\cite{PDG03}:
\begin{equation}
\label{eq:modind}
{\mathcal{L}}^{e-N}=\frac{G_F}{\sqrt{2}}
\sum_q [C_{1q}
{\bar e}\gamma_\mu
    \gamma^5 e {\bar q}\gamma^\mu q
+C_{2q} {\bar e} \gamma_{\mu} e {\bar q} \gamma^{\mu} \gamma^{5} q],
\end{equation}
where $C_{1q}$ are the vector quark couplings and $C_{2q}$ are
the axial-vector quark couplings.  The $Q_{weak}$ experiment will
set significant new constraints on $C_{1u}$ and $C_{1d}$ as shown in
Figure~\ref{fig:c1_u_d}.  The shown reduction in the size of the allowed
ellipse would result from the precision of the $Q_{weak}$ measurement and
its complementarity to existing data. To indicate the ability of
low energy experiments to probe new physics possibilities at high
energy scales, we consider the case of the vector couplings and define
the following~\cite{Ram00}:
\begin{equation}
{\mathcal{L}}^{PV}_{NEW}=\frac{g^2}{4\Lambda^2} {\bar e}
    \gamma_\mu\gamma_5 e\sum_q h_V^q\ {\bar q}
    \gamma^\mu q,
\end{equation}
as a particular parameterization of new physics.  Here, $g$, $\Lambda$, and
$h_V^q$ are, respectively, the coupling constant, the mass scale,
and effective coefficients that are used to parameterize
possible new physics beyond the Standard Model.
In the case of $Q_{weak}$,
a 4\% measurement of $Q_{W}^{p}$ would test new physics
scales up to
\begin{equation}
\frac{\Lambda}{g} \sim \frac{1}{\sqrt{\sqrt{2} G_F |\Delta Q_{W}^{p}|}}
\approx 4.6~\hbox{TeV}.
\end{equation}
Thus, precision
low energy experiments with virtual exchange probes are sensitive to
new physics at energies comparable to
the scales where colliders can produce the new physics-related
particles directly.

\subsubsection{Parity Violation in Deep Inelastic Scattering}
\label{sec:disparity}

The original experiment of Prescott~{\it et al.}~\cite{Pre79} utilized parity-violating
deeply inelastic (DIS) electron scattering
from deuterium.  Recently, there have been several proposals
to measure this process with higher precision.  One proposed experiment would
use 36 and 39 GeV polarized electron beams at SLAC~\cite{Bos03}.  Experiments
have also be proposed at Jefferson Lab with 6 GeV~\cite{Zhe03} and with the
future upgraded energy of 12 GeV~\cite{CDR03}.

The parity-violating asymmetry for scattering longitudinally polarized electrons from partons in an
unpolarized deuterium target is given by~\cite{Cah78}:
\begin{equation}
\label{eq:a_dis}
A_d = \frac{\sigma_R-\sigma_L}{\sigma_R+\sigma_L} =\left(\frac{3 G_F Q^2}{\pi \alpha 2
\sqrt{2}}\right) \frac{2 C_{1u}-C_{1d}[1+R_s(x)]+Y(2C_{2u}-C_{2d})R_v(x)} {5 +R_s(x)},
\end{equation}
where $C_{1u}, C_{1d}, C_{2u},\ \hbox{and}\ C_{2d}$ are the model independent electron quark couplings defined
in Equation~\ref{eq:modind}.  The ratios $R_s(x)$ and $R_v(x)$ are ratios of quark distribution functions.  The
quantity $Y$ is a function of the kinematic variable $y=\nu/E$, where $\nu=E-E'$ is the energy lost by an
incident electron of energy $E$ scattering to an electron of energy $E'$.  A simplified expression for the
asymmetry, appropriate to the kinematics of the proposed measurements, is:
\begin{equation}
A_d \sim - 10^{-4} Q^2 [(2C_{1u} - C_{1d}) + Y(2 C_{2u} - C_{2d})]
\end{equation}
The dependence of this expression on $Y$ allows for separate extraction
of the vector ($2C_{1u}-C_{1d}$) and axial-vector ($2C_{2u}-C_{2d}$)
combinations.  For typical kinematics of the proposed experiments,
$A \sim 100\ \hbox{ppm}$, which is a large asymmetry compared to the
current generation of experiments.
The experiments can be interpreted in terms of the Standard
Model by using the Standard Model values of the weak charges from
Table~\ref{tab:charges}:
\begin{eqnarray}
C_{1u} &=& g^e_A g^u_V = -\frac{1}{2} + \frac{4}{3} \sin^2 \theta_W \nonumber \\
C_{1d} &=& g^e_A g^d_V = \frac{1}{2} - \frac{2}{3} \sin^2 \theta_W \nonumber \\
C_{2u} &=& g^e_V g^u_A = -\frac{1}{2} + 2 \sin^2 \theta_W \nonumber \\
C_{2d} &=& g^e_V g^d_A = \frac{1}{2} - 2 \sin^2 \theta_W. \nonumber \\
\end{eqnarray}
The currently proposed DIS-parity experiments aim to provide precision
measurements of $\sin^2 \theta_W$ in the range $Q^2 \sim 2 - 20\ \hbox{GeV}^2$.
As seen in Figure~\ref{fig:runs2w}, this is the same $Q^2$ range
in which the NuTeV collaboration reports a deviation from the Standard
Model expectation.  As pointed out in section~\ref{sec:qweak}, the NuTeV
experiment could potentially be explained by more conventional mechanisms.
The DIS-parity experiments on deuterium would be an important cross-check
on this result, because they would not be impacted by the same uncertainties
in nuclear effects and nuclear parton distributions that are present
in the NuTeV measurement on an iron target.  When analyzed in the model
independent framework described in section~\ref{sec:qweak},  the
primary impact of the
DIS-parity experiments described here will be to reduce the allowed
phase space of the axial coupling combination, $2C_{2u}-C_{2d}$ .  The
current limits on these couplings
are shown in Figure~\ref{fig:c2u_c2d}.  Possible new physics extensions
to the Standard Model that would be present in DIS-parity experiments
have been considered in Kurylov, {\it et al.}~\cite{Kur04}.

The proposed DIS-parity experiment at SLAC~\cite{Bos03} would scatter
36 and 39 GeV polarized electron beams from a 100 cm liquid deuterium
target.  Scattered electrons would be detected in two large solid
angle magnetic spectrometers at a mean scattering angle of $\sim 12^{\circ}$.
The electrons would be identified and counted in an array of Pb-glass blocks,
with a gas \v Cerenkov counter to aid in particle
identification.  The experiment will be run at two different beam energies
because the electron spin precesses by $\pi$ between the two beam
energies due to its anomalous moment.  The resulting change in sign
of the measured asymmetry is a useful systematic check for false asymmetries.
This anticipated result of this experiment would be a
0.4\% (combined relative error) determination of $\sin^2 \theta_W$ at
$Q^2 = 20\ \hbox{GeV}^2$.

At Jefferson Lab, there are ideas to do DIS-parity at both 6 GeV and eventually
the upgraded 12 GeV beam energy.  The proposal at 6 GeV~\cite{Zhe03}
would use a 140 $\mu$A, 6 GeV, 80\% polarized electron beam incident on
a 15 cm liquid deuterium target.  The scattered electrons will be detected
in the two Hall A High Resolution Spectrometers (HRS) at a scattering
angle of $19^{\circ}$.  The expected asymmetry in this kinematics
is about 131~ppm and the anticipated result would be a
1.0\% (combined relative error) determination of $\sin^2 \theta_W$ at
$Q^2 = 2\ \hbox{GeV}^2$.  With an upgraded
CEBAF accelerator energy of 12 GeV, a possible 11 GeV
DIS-parity experiment have been evaluated~\cite{CDR03}.  An experiment
with 90 $\mu$A of 80\% polarized electron beam incident
on a 60 cm liquid deuterium
target with upgraded Hall A or Hall C spectrometers could give
a 0.6\% (combined relative error) determination of
$\sin^2 \theta_W$ at $Q^2 = 3\ \hbox{GeV}^2$.
One potential concern with the Jefferson Lab measurements is the presence of higher
twist QCD contributions at the relatively low momentum transfer involved.
The expression for the asymmetry in Equation~\ref{eq:a_dis} only includes
twist-two ("free-parton") contributions.  The formalism for including
higher twist (quark-gluon correlation) contributions is described in
~\cite{Cas85}. If higher twist effects turn out to be significant, then measurements at
several values of $Q^2$ and $x$ will be necessary to separate the higher
twist effects from the weak mixing angle determination.

\subsection{Summary}

Since the earliest measurements of Prescott and collaborators~\cite{Pre79},
basic techniques of parity-violating electron scattering have remained
more or less the same. But the achievable precision has greatly improved
as a result of new high intensity electron beams, advances in polarized
beam technology, and technical advances in the feedback and laser systems
that are needed for polarized beam delivery. In the last decade, the focus of
new experiments has been to use the neutral weak interaction as a probe of
hadron structure in a way that is complementary to the large body of
information from electromagnetic probes. Although the experiments remain challenging,
results from the first round of experiments are now available.  While these data
have provided an initial look at the role of strange quarks in the proton's
electromagnetic properties, ambiguities remain because these experiments cannot yet distinguish
between small values of $G_M^s$ and $G_E^s$ or cancellations due to differing signs.
In the next few years, new results will be available from HAPPEX, PVA4 and from
the G0 Experiment at Jefferson Lab, all of which should help answer this question.
The precision with which this generation of experiments will constrain
these hadronic structure issues will be sufficient to allow for a future program
of parity-violating electron-nucleon scattering that is again focused on a determination
$\sin^2\theta_W$.

The authors gratefully acknowledge financial support from the
National Science Foundation. We also wish to acknowledge all members
of the SAMPLE collaboration, who are listed in Refs.~\cite{Ito03} and~\cite{Spa03}.
Much of the work discussed in this paper has been supported both by the NSF and
the U.S.~Dept.~of Energy.


\end{document}